\documentclass[3p,preprint]{elsarticle}

\usepackage{lineno,hyperref}
\usepackage{bm}
\usepackage{placeins}
\usepackage{array}
\usepackage{caption}

\usepackage{booktabs,makecell,multirow,array,colortbl,dcolumn,stfloats} 

\usepackage{framed} 
\usepackage{multicol} 
\usepackage{nomencl} 
\makenomenclature
\renewcommand*\nompreamble{\begin{multicols}{2}}
\renewcommand*\nompostamble{\end{multicols}}

\modulolinenumbers[5]

\usepackage{color}
\usepackage{relsize}
\usepackage[utf8]{inputenc}
\usepackage{mathtools}
\usepackage{amsmath}
\usepackage{amssymb}
\usepackage[size=small]{todonotes} 
\usepackage{tcolorbox} 
\usepackage{pdflscape} 
\usepackage{geometry} 
\usepackage{setspace}\usepackage{threeparttable} 
\usepackage{pdfpages}
\usepackage{lipsum}
\newcommand{\ve}[1]{\boldsymbol{#1}} 
\newcommand{\te}[1]{\boldsymbol #1} 

\newcommand{\diffp}[2]{\frac{\partial #1}{\partial #2}} 
\newcommand{\diffv}[2]{\frac{\delta #1}{\delta #2}} 
\newcommand{\inte}[3]{\mathop{\int}_{ #1} #2 \; \mathrm{d} #3} 
\newcommand{\intee}[4]{\mathop{\int}_{ #1}^{ #2} #3 \; \mathrm{d} #4} 

\newcommand{\dev}{\mathrm{dev}}

\newcommand{\calE}{{\cal E}}
\newcommand{\calW}{{\cal W}}
\newcommand{\calB}{{\cal B}}

\newcommand{\calH}{{\cal H}}
\newcommand{\calF}{{\cal F}}

\newcommand{\Gc}{G_\mathrm{c}}
\newcommand{\sig}{\te{\sigma}}

\newcommand{\sigy}{\sigma^\mathrm{y}}
\newcommand{\eps}{\te{\varepsilon}}
\newcommand{\epse}{\te{\varepsilon}^\mathrm{e}}
\newcommand{\epsp}{\te{\varepsilon}^\mathrm{p}}

\newcommand{\psie}{\psi^\mathrm{e}}
\newcommand{\psip}{\psi^\mathrm{p}}

\newcommand{\qal}{\te{q}_\alpha}
\newcommand{\gfat}{\tilde{g}}

\newcommand{\Def}{\coloneqq }

\journal{Engineering Fracture Mechanics}

\begin{document}

\begin{frontmatter}

\title{Overview of phase-field models for fatigue fracture in a unified framework}


\author[mymainaddress]{Martha Kalina}

\author[mymainaddress]{Tom Schneider}

\author[mymainaddress]{Jörg Brummund}

\author[mymainaddress,mythirdaddress]{Markus K\"{a}stner\corref{mycorrespondingauthor}}
\cortext[mycorrespondingauthor]{Corresponding author}
\ead{markus.kaestner@tu-dresden.de}

\address[mymainaddress]{Chair of Computational and Experimental Solid Mechanics, TU Dresden, Dresden, Germany}
\address[mythirdaddress]{Dresden Center for Computational Materials Science (DCMS), TU Dresden, Dresden, Germany}

\begin{abstract}	
The phase-field method has gained much attention as a novel method to simulate fracture due to its straightforward way allowing to cover crack initiation and propagation without additional conditions. More recently, it has also been applied to fatigue fracture due to cyclic loading. This publication gives an overview of the main phase-field fatigue models published to date. {For the first time, we present all models in a unified variational framework for best comparability}. Subsequently, the models are compared regarding their most important features. It becomes apparent that they can be classified in mainly two categories according to the way fatigue is implemented in the model -- that is as a gradual degradation of the fracture toughness or as an additional term in the crack driving force. We aim to provide a helpful guide for choosing the appropriate model for different applications and for developing existing models further.
\end{abstract}


\end{frontmatter}

\section{Introduction} 

Fatigue fracture is the main cause of failure in engineering structures \cite{stephens_metal_2000}. A fatigue crack usually undergoes three stages \cite{bathias_fatigue_2010}: The crack initiation stage, followed by stable crack propagation and sudden residual fracture. For many engineering components, the structure is designed to withstand crack initiation, e.\,g. with the help of component S-N curves (also called \textsc{Wöhler} curves). But especially in thin-walled parts, the resistance against fatigue crack growth can be decisive for the design process as well. {Often, \textsc{Paris} curves \cite{paris_critical_1963}, which describe the fatigue crack growth rates in the material, are used to estimate the crack growth for a given number of load cycles, e.\,g. within one inspection interval. However, traditional methods of fracture mechanics are limited to straight cracks and known crack paths. More advanced numerical techniques for the estimation of crack growth are currently under development.}
{ Modelling fracture using sharp crack representation comes with certain drawbacks. 
E.\,g. cohesive zone models \cite{kuna_numerische_2010} suffer from the problem of describing the topology of evolving cracks and also require a predefined crack path. The XFEM method \cite{moes_finite_1999}, on the other hand, uses enriched shape functions in order to capture the crack, which, at the latest in 3D, becomes very complex in order to cover all possible crack patterns within an element.}
From this perspective, the phase-field method for fracture is advantageous as it describes the crack topology with an additional field variable. The emerging coupled problem covers crack initiation, deflection, branching and merging of cracks in a straightforward way. Due to its flexibility, this method has gained attention and advancement in the past ten years.

After the pioneering works of \textsc{Francfort} and \textsc{Marigo} \cite{francfort_revisiting_1998} and \textsc{Bourdin} et al. \cite{bourdin_numerical_2000,bourdin_variational_2008} regarding the variational formulation of fracture and the regularisation of the crack geometry, as well as  \textsc{Miehe} et al. \cite{miehe_thermodynamically_2010,miehe_phase_2010} regarding  model formulation and implementation, a variety of different approaches to phase-field modelling of static brittle fracture have been published, see \cite{ambati_review_2015} for an overview. The various extensions to ductile fracture are reviewed in \cite{alessi_comparison_2018}, see furthermore \cite{dammass_unified_2021} for an overview of viscous phase-field models. More recently, fatigue fracture has also been a topic of intensive research in the phase-field community. It is the aim of this work to give an overview of the models, explain differences and highlight the loading types and scope of application they might be suitable for. 

When discussing the modelling of fatigue cracks it is important to consider the different mechanisms that lead to fracture, depending on material and loading type.  Under small loading amplitudes, material can withstand large numbers of load cycles (High cycle fatigue -- HCF). The material behaves macroscopically mostly elastic. On the other hand, in low cycle fatigue (LCF), load amplitudes are higher, leading to significant inelastic effects, especially around the crack tip. 
The transition between HCF to LCF depends on the material. For metals, $10^2$ to $10^4$ load cycles are considered to be LCF \cite{bathias_fatigue_2010}. LCF cracks are correlated best with elastic-plastic strain quantities while HCF cracks are mostly stress-controlled \cite{bathias_fatigue_2010}.
Furthermore, not only the load amplitude, but also the mean load, the multiaxiality of the loading and the crack opening mode \cite{radaj_ermudungsfestigkeit_2007}  can have significant influence on fatigue life. The same applies to crack closure effects caused by plastic deformation and roughness of the crack flanks, among others \cite{vasudeven_review_1994}.

Historically, due to their great industrial relevance, metals are the materials studied best regarding their fatigue behaviour. Fatigue in metals arises from plasticity \cite{bathias_fatigue_2010}. For LCF, macroscopic plastic deformations accompany the crack. However, even for macroscopic stresses below the elastic limit -- typical for HCF -- stress concentrations at defects on the grain-scale occur, which lead to plastic microdeformations \cite{lemaitre_mechanics_1998}. This effect causes cyclic work-hardening or softening of the material, i.\,e. increasing or decreasing stress amplitudes in a strain-controlled experiment, compared to the monotonic stress-strain curve \cite{bathias_fatigue_2010}. 

Crack initiation in metals is caused by dislocations in the polycrystalline material. These dislocations accumulate in permanent slip bands, driven by shear stress components and finally lead to material separation \cite{bathias_fatigue_2010}. Slip bands often form at stress concentrations, e.\,g. at notches, imperfections, voids and inclusions \cite{radaj_ermudungsfestigkeit_2007}. Merging of the, at first, microscropic cracks finally leads to macroscopic crack initiation. This initiation phase can take up to ninety percent of the component's fatigue life \cite{bathias_fatigue_2010}. Afterwards, the crack evolves into a so-called \textit{long crack}, i.\,e. visible crack, with alternating plastic slips on each flank \cite{lemaitre_mechanics_1998} which is well-described by \textsc{Paris} law \cite{rovinelli_assessing_2017} and then finally undergoes sudden residual fracture. 

Fatigue in polymers, on the other hand, is mainly caused by formation of cavities and cavitations. Macromolecules are degraded progressively. Although the mechanisms leading to fatigue in polymers are manifold and strongly depend on the type of polymer, damage is mostly controlled by shear, principal strains and the hydrostatic part of the stress tensor. In contrast to metals, cracks can evolve under compressive or hydrostatic stress. \cite{bathias_fatigue_2010} 
Especially elastomers call for the use of finite strain measures even in fatigue simulations as well as rate-dependent models.

The majority of the phase-field fatigue models mentioned in this overview is either meant for or at least applied to metals, yet there are also a few for other material classes. The different models mainly vary concerning their fatigue variable, which describes the cyclic loading history of the material, and the way this fatigue variable is incorporated into the model. With regard to the latter, this paper identifies two main model classes most phase-field fatigue models fit into: 
Those with degraded facture toughness (\textit{type A}) and those with additional crack driving force (\textit{type B}). { In contrast to the review \cite{cui_applications_2023},} this paper presents all models in a unified framework to allow for better comparability and to discuss common features and differences. This ought to be a helpful basis for further development of phase-field models for cyclic loads. Furthermore it is meant as a guide for choosing a model for a component of a certain material undergoing a specific loading type. Further demands regarding the simulation time or physical rigour of the model may also to be taken into consideration.

The paper is structured as follows. Section 2 outlines a general framework for phase-field fatigue models which comprises most models presented later. A variational formulation is used. In addition, a short overview of other derivation strategies is given subsequently. Section 3 includes a short description of all mentioned models as well as a table listing model features for clarity. Section 4 discusses the main model features. The characteristics of model type \textit{A} and \textit{B} (see above) are emphasised by a numerical example. The paper terminates with conclusion and outlook.

\begin{table*}[!t]
	\begin{framed}
		\nomenclature{$\calB $}{Domain }
		\nomenclature{$\partial\calB $}{ Boundary}
		\nomenclature{$\ve{x} $}{Location }
		\nomenclature{$t, \tau $}{Time }
		\nomenclature{$\eps $}{Total strain }
		\nomenclature{$\epse $}{Elastic strain }
		\nomenclature{$\epsp $}{Plastic strain }
		\nomenclature{$\ve{u} $}{Displacement }
		\nomenclature{$\te{\alpha} $}{Kinematic hardening }
		\nomenclature{$\alpha $}{Isotropic hardening }
		\nomenclature{$\ve{q}_\alpha $}{Set of plastic variables }
		\nomenclature{$d $}{Fradcture phase-field }
		\nomenclature{$\calF $}{Fatigue variable }
		\nomenclature{$W $}{Generating energy density functional}
		\nomenclature{$\psi $}{Free energy density }
		\nomenclature{$\Delta $}{Dissipative part of $W$ }
		\nomenclature{$\sig $}{Stress }
		\nomenclature{$\te{\chi} $}{Backstress for kinematic hardening }
		\nomenclature{$p $}{Stress associated with isotropic hardening}
		\nomenclature{$W_\mathrm{el} $}{Elastic part of $W$ }
		\nomenclature{$W_\mathrm{pl} $}{Plastic part of $W$ }
		\nomenclature{$W_\mathrm{frac} $}{Fracture part of $W$ }
		\nomenclature{$W_\mathrm{fat} $}{Fatigue part of $W$ }
		\nomenclature{$W_\mathrm{reg} $}{Regularisation part of $W$ }
		\nomenclature{$g $}{Degradation function }
		\nomenclature{$\psie_+ $}{Tensile part of elastic energy density }
		\nomenclature{$\psie_- $}{Compressive part of elastic energy density }
		\nomenclature{$\psip $}{Energy density of hardening}
		\nomenclature{$\Delta^\mathrm{p} $}{Energy density of plastic dissipation }
		\nomenclature{$\phi^\mathrm{p} $}{Plastic dissipation potential }
		\nomenclature{$\phi^\mathrm{visc} $}{Viscous dissipation potential }
		\nomenclature{$\sig^\mathrm{ov} $}{Overstress }
		\nomenclature{$\te{\varPhi} $}{Damper strain }
		\nomenclature{$\Gc $}{Fracture toughness }
		\nomenclature{$\gamma $}{Regularised crack surface energy density}
		\nomenclature{$\omega $}{Local part of $\gamma$ }
		\nomenclature{$c_{\omega} $}{Constant of $\gamma$ }
		\nomenclature{$\ell $}{Regularisation length }
		\nomenclature{$\Delta^\mathrm{reg} $}{Regularisation part of $W$ }
		\nomenclature{$\phi^\mathrm{reg} $}{Energy density of regularisation }
		\nomenclature{$\eta $}{Viscous regularisation constant }
		\nomenclature{$W^A_\mathrm{fat} $}{$W_\mathrm{fat}$ for model version "A"}
		\nomenclature{$W^B_\mathrm{fat} $}{$W_\mathrm{fat}$ for model version "B"}
		\nomenclature{$h $}{Fatigue degradation function of model version "A" }
		\nomenclature{$\tilde{g} $}{Degradation function of $W^B_\mathrm{fat} $}
		\nomenclature{$H $}{Fatigue function of model version "B" }
		\nomenclature{$\calE $}{Generating functional }
		\nomenclature{$\calE_\mathrm{ext} $}{Work of external forces }
		\nomenclature{$\bar{f} $}{Volume force }
		\nomenclature{$\bar{\tau} $}{Traction vector }
		\nomenclature{$\partial\calB^\mathrm{N} $}{\textsc{Neumann} boundary }
		\nomenclature{$\partial\calB^\mathrm{D} $}{\textsc{Dirichlet} boundary }
		\nomenclature{$\Pi^\tau $}{Incremental rate form of $\calE$ }
		\nomenclature{$t_n $}{Last timestep }
		\nomenclature{$\calW $}{Conditions for variational principle }
		\nomenclature{$d_n $}{Phase-field of last timestep }
		\nomenclature{$\bar{u} $}{Displacement boundary conditional }
		\nomenclature{$\ve{n} $}{Normal vector }
		\nomenclature{$a,b,q,\kappa $}{Material constants }
		\nomenclature{$\sigy $}{Yield stress }
		\nomenclature{$f^\mathrm{p} $}{Plastic yield function }
		\nomenclature{$\lambda $}{Plastic multiplier }
		\nomenclature{$\calH $}{History variable of crack driving force }
		\nomenclature{$f^d $}{Yield condition of phase-field problem }
		\nomenclature{$\kappa $}{Fatigue degradation parameter }
		\nomenclature{$\sig^\mathrm{eq} $}{Equilibrium stress }
		\nomenclature{$\lambda^\infty $}{Penalty parameter }
		\nomenclature{$\sig^* $}{Undamaged stress }
		\nomenclature{$\sig_+ $}{Tensile stress part }
		\nomenclature{$\sig_- $}{Compressive stress part }
		\nomenclature{$w $}{Measurement of CT specimen }
		\nomenclature{$a_0 $}{Initial crack length }
		\nomenclature{$F $}{Load (force) }
		\nomenclature{$R $}{Load ratio }
		\nomenclature{$G $}{Energy release rate }
		\printnomenclature
	\end{framed}
\end{table*}

\clearpage

\section{General framework for phase-field fatigue models}

The respective models are compared using a general phase-field framework for fatigue fracture outlined in the following. Besides the way of integrating fatigue, the models cover a variety of modelling features including various types of plasticity and, albeit few of them, viscous behaviour. At first, the derivation of the governing equations in this chapter is limited to  elastic-plastic cyclic behaviour, an alternative for viscous behaviour is given later. However, since most models use viscous regularisation for numerical reasons, it is included standardly. Some other deviations from the general derivation presented here occur for a few models and will become clear  in Section \ref{sec:overview}. 
Nomenclature from the original papers is commonly abandoned for the sake of comparability.  The way of derivation and nomenclature partly follow \cite{noii_bayesian_2021a} and \cite{alessi_comparison_2018}, though not strictly.
The modelling framework is presented using a variational framework. Still, a brief overview of other ways of derivation is given at the end of the section.

\subsection{Model derivation via variational framework}
\label{sec:framework}
The domain under consideration is $\calB \subset \mathbb{R}^n$ with its boundary $\partial\calB$ and material points described by location $\ve{x}$ at time $t$. In a small strain setting, the total strain $\eps(\ve{x},t)$ can be decomposed additively into elastic strain $\epse(\ve{x},t)$ and plastic strain $\epsp(\ve{x},t)$ 
\begin{equation}
	\eps \Def  \frac{1}{2} \left( \nabla\ve{u} + \nabla\ve{u}^\top \right) = \epse + \epsp
\end{equation} 
with $\ve{u}(\ve{x},t)$ being the displacement. Plastic deformations can lead to hardening, which is described by the kinematic and isotropic hardening variables $\te{\alpha}(\ve{x},t)$ and  $\alpha(\ve{x},t)$, respectively. The plastic variables are summarised in the set $\qal = \left\{ \epsp,\te{\alpha},\alpha \right\}$. Cracks are described in a regularised manner using the phase-field variable $d(\ve{x},t)$, with intact material being marked by $d=0$ and fully fractured material marked by $d=1$. The cyclic loading and damage history is described by a scalar fatigue variable $\mathcal{F}(\ve{x},t)$.
Dependencies on space, time and other variables are omitted hereafter, if not particularly necessary.

\subsubsection*{Energy functional}

In order to set up a variational principle later on, a generating functional of energy density type 
\begin{equation}
	W(\eps,d,\nabla d,\dot{d},\qal,\dot{\te{q}}_{\alpha};\calF) \Def \psi(\eps,d,\qal) + \Delta(\eps,d,\nabla d,\dot{d},\qal,\dot{\te{q}}_{\alpha};\calF)
\end{equation}
is defined which consists of a free energy density $\psi$ and a dissipative part $\Delta$. From the \textsc{Clausius-Duhem} inequality 
\begin{equation}
		\sig:\dot{\eps} - \diffp{\psi}{\eps}:\dot{\eps} - \diffp{\psi}{\qal}:\dot{\te{q}}_{\alpha} - \diffp{\psi}{d}\,\dot{d} \geq 0
\end{equation}
we can identify 
\begin{equation}
	-\diffp{\psi}{\epsp} = \diffp{\psi}{\eps} =:  \sig \quad -\diffp{\psi}{\te{\alpha}}  =: \te{\chi} \quad -\diffp{\psi}{\alpha} =: p 
\end{equation}
the stress $\sig$, a backstress tensor $\te{\chi}$ for kinematic hardening and the stress-like quantity $p$ associated with isotropic hardening.
For clarity, the generating density functional $W$ is here decomposed into
\begin{equation}
W \Def W_{\mathrm{el}}(\epse,d) + W_{\mathrm{pl}}(\eps,d,\qal,\dot{\te{q}}_{\alpha}) + W_\mathrm{frac}(d,\nabla d) 
+ W_\mathrm{fat}(d,\nabla d;\calF) + W_\mathrm{reg}(\dot{d})
\end{equation} 
the elastic free energy $W_\mathrm{el}$, the plastic part $W_\mathrm{pl}$, the contributions from fracture $W_\mathrm{frac}$ and fatigue $W_\mathrm{fat}$, respectively, and the viscous regularisation $W_\mathrm{reg}$. 
The \textbf{elastic} energy density 
\begin{equation}
W_\mathrm{el}(\epse,d) \Def g(d)\, \psie_+(\epse) + \psie_-(\epse)
\end{equation}
consists of a degraded part (often the tensile part) $\psie_+$ with the degradation function $g(d)$ and a part (often the compressive part) $\psie_-$, which remains undegraded. For this split, various concepts are used by the  models compared here, the most common one being the split by \textsc{Amor} et al. \cite{amor_regularized_2009}.
For the stress\footnote{Some models use a different stress definition, see Section \ref{sec:stress}.} it follows
\begin{equation}
	\sig(\epse) \Def \diffp{W_\mathrm{el}}{\epse} = g(d) \sig_+(\epse) + \sig_-(\epse)
\end{equation}
while the (virtually) undamaged stress is 
\begin{equation}
	\sig^*(\epse) \Def \sig_+ + \sig_-.
\end{equation}
The energy density related to \textbf{plasticity}\footnote{Some models also have dependencies on $\nabla \alpha$ in case of gradient plasticity or an explicit strain measure for ratchetting, e.\,g. \textsc{Ulloa} et al. \cite{ulloa_phasefield_2021}.} 
\begin{equation}
W_\mathrm{pl}(\eps,d,\qal,\dot{\te{q}}_{\alpha}) \Def g(d) \psip(\eps,\qal) + g(d) \Delta^\mathrm{p}(\eps,d,\qal,\dot{\te{q}}_{\alpha})
\end{equation}  
consists of a hardening contribution $\psip$ and a dissipative contribution $\Delta^\mathrm{p}$
\begin{equation}
	\Delta^\mathrm{p}(\eps,d,\qal,\dot{\te{q}}_{\alpha}) = \intee{0}{t}{\phi^\mathrm{p}(\eps,d,\qal,\dot{\te{q}}_{\alpha})}{\tau} 
\end{equation}
which follows from a plastic dissipation potential $\phi^\mathrm{p}$. Usually, but not always, both are degraded by the same degradation function $g(d)$ as the elastic contribution. The dissipation potential can e.\,g. be derived from the principle of maximum dissipation. 
\paragraph{Remark}
In order to create an explicitly viscous model (such as in \textsc{Loew} et al. \cite{loew_fatigue_2020,loew_accelerating_2020}), $\psi^\mathrm{p}$ and $\phi^\mathrm{p}$ can be substituted by their viscous counterparts, e.\,g.
\begin{equation}
	\psi^\mathrm{visc} = \intee{0}{t}{\sig^\mathrm{ov}:\dot{\eps}}{\tau} \quad \text{and} \quad \phi^\mathrm{visc} = \sig^\mathrm{ov}:\dot{\te{\varPhi}}
\end{equation}
with the non-equilibrium stress $\sig^\mathrm{ov}$ and the inelastic variable set now including the viscous strain $\qal = \te{\varPhi}$.


The damage dissipation density due to formation of \textbf{crack surface} is given by
\begin{equation}
W_\mathrm{frac}(d,\nabla d)  \Def \Gc \gamma(d,\nabla d) 
\end{equation}
wherein $\Gc$ is the fracture toughness and the regularised crack surface density $\gamma$ is
\begin{equation}
	\gamma(d,\nabla d)  \Def \frac{1}{c_\omega}\left( \frac{\omega(d)}{\ell} + \ell\nabla d\cdot\nabla d \right).
\end{equation}
For the latter, the two most common formulations are so-called \textsc{Ambrosio-Tortorelli} \cite{ambrosio_approximation_1990} (AT) 1 with $c_\omega=\frac{3}{8},\omega(d)=d$ and AT 2 with $c_\omega=\frac{1}{2},\omega(d)=d^2$.
See \cite{delorenzis_numerical_2020} and the literature cited therein for possible other choices for the local part of the dissipated fracture energy density $w(d)$.
The viscous \textbf{regularisation} term
\begin{equation}
	W_\mathrm{reg} = \Delta^\mathrm{reg} = \intee{0}{t}{\phi^\mathrm{reg}(\dot{d})}{\tau} \quad \text{with} \quad \phi^\mathrm{reg} = \frac{1}{2}\eta\dot{d}^2
\end{equation}
ensures numerical stability in cases of rapidly evolving cracks. 
Finally, for the \textbf{fatigue} contribution $W_\mathrm{fat}$, most models studied in this paper\footnote{Except for \textsc{Aygün} et al. \cite{aygun_coupling_2021a} and \textsc{Lo} et al. \cite{lo_phasefield_2019}, see Section \ref{sec:overview}} use one of the two structures 
\begin{equation}
{\color{blue}W_\mathrm{fat}^A(d,\nabla d;\calF) \Def \left(h(\calF)-1\right)\frac{\Gc}{c_w}\left(\frac{w(d)}{\ell}+\ell\nabla d\cdot\nabla d\right)} \quad \text{or} \quad {\color{red}W_\mathrm{fat}^B(d;\calF) \Def \gfat(d) H(\calF) }.
\end{equation}
They will be called $A$-models and $B$-models hereafter. The following derivations include both fatigue contribution terms at once, which are marked by the colours {\color{blue} blue (A)} and {\color{red}red (B)} for distinguishability.\footnote{Please refer to the online edition of the paper for a coloured version.} 
The fatigue degradation function $h(\calF)$ should be scalar and without unit whereas the additive fatigue contribution $H(\calF)$ should be an energetic quantity. This becomes clear when recapitulating the generating density functional \footnote{Some prefer to write $h(\calF)\Gc\gamma$ in rate form due to process dependency of quantity, e.\,g. \cite{carrara_framework_2019,alessi_phenomenological_2018}.} with both contributions
\begin{equation}
	W = \underbrace{g(d)\, \psie_+ + \psie_- + g(d) \psip}_{\Def\psi} + \underbrace{ g(d) \Delta^\mathrm{p} +  {\color{blue}h(\calF)}\Gc\gamma + {\color{red} \gfat(d)H(\calF)} + \Delta^\mathrm{reg}}_{\Def \Delta}.
\end{equation}
Obviously, in type $A$ models, the fracture toughness is reduced gradually in order to model the decreasing resistance of the material to withstand cracks due to cycling loading. Thereby, the pseudo energy density $W$ is reduced. On the other hand, $B$-type models yield   an additional energy term which increases the crack driving force, as will be shown later on.\footnote{Some models ascribe more parts of $W$ to the free energy density $\psi$, leading to additional stress terms. See Section \ref{sec:stress}.} At the same time, this also increases the total pseudo energy. The coefficient $\gfat(d)$ of the additive fatigue contribution $H(\calF)$ can be, but is not always equal to $g(d)$.





\subsubsection*{Variational principle}


Variational analysis is used to derive the model equations. The generating density functional $W$ is integrated to form the generating functional 
\begin{equation} \label{eq:enfun}
\calE(\eps,d,\nabla d,\dot{d},\qal,\dot{\te{q}}_{\alpha};\calF) \Def \inte{\mathcal{B}}{W(\eps,d,\nabla d,\dot{d},\qal,\dot{\te{q}}_{\alpha};\calF)}{v} - \calE_\mathrm{ext}(\ve{u}) ,
\end{equation}
also considering the work from external forces $\calE_\mathrm{ext}$ due to volume force  $\bar{f}$ and traction vector  $\bar{\tau}$ 
\begin{equation}
\calE_\mathrm{ext}(\ve{u}) \Def \inte{\mathcal{B}}{\bar{f}\cdot\ve{u}}{v} + \inte{\partial\mathcal{B}^\mathrm{N}}{\bar{\tau}\cdot\ve{u}}{a}.
\end{equation}
The generating functional is now formulated in an incremental form $\Pi^\tau$ for the  time step $ t-t_n$ 
\begin{align} \label{eq:pitau}
\Pi^\tau(\eps,d,\nabla d,\dot{d},\qal,\dot{\te{q}}_{\alpha};\calF) \Def & \, \calE(t) - \calE (t_n) \nonumber \\
= & \int_{\mathcal{B}} \Biggl\{  \psi(t)-\psi(t_n) + {\color{blue}h(\calF)}\Gc \left(\gamma(t)-\gamma(t_n)\right) + {\color{red}\left(\tilde{g}(t)-\tilde{g}(t_n)\right)H(\calF)} \biggr. \nonumber \\ & + \left. \intee{t_n}{t}{\left[\phi^\mathrm{p}(\tau) + \phi^\mathrm{reg}(\tau)\right]}{\tau}  - \left(\bar{f}\cdot\ve{u}(t) - \bar{f}\cdot\ve{u}(t_n)\right) \right\}\,\mathrm{d}v - \inte{\partial\mathcal{B}^\mathrm{N}}{\left\{\bar{\tau}\cdot\ve{u}(t)-\bar{\tau}\cdot\ve{u}(t_n)\right\}}{a} .
\end{align}The fatigue variable $\calF$ is assumed to be constant for the moment considered here (i.\,e. within an increment) since due to its nature, $\calF$ changes on a much larger time scale than e.\,g. $\eps$ or $\qal$, which are subject to oscillation over the cycles.
The incremental variational principle reads 
\begin{equation}
\{ \ve{u},d,\qal \} = \arg \left\{ \min_{\ve{u}\in\calW_{\bar{u}}} \, \min_{d\in\calW_{d_n}} \, \min_{ \qal \in \calW_{p}} \Pi^\tau(\eps,d,\nabla d,\dot{d},\qal,\dot{\te{q}}_{\alpha};\calF) \right\}
\end{equation}
with the spaces of admissible functions, including conditions for the displacement on the boundaries and  irreversibility of the phase-field
\begin{align}
\calW_{\bar{u}} &\Def \{ \ve{u}\in \mathbb{R}^3 \,|\, \ve{u}=\bar{\ve{u}} \text{ on } \partial\mathcal{B}^\mathrm{D} \}\\
\calW_{d_n} &\Def \{ d\in \mathbb{R} \,|\, d\geq d_n \} \\
\calW_{p} &\Def \{ \qal\in \mathbb{R}^n \}.
\end{align}
Next, stationarity conditions for the displacement, the plastic variables and the phase-field are exploited one by one in order to derive the model equations. 

\subsubsection*{Displacement}

The variational derivative $\delta_u$ of $\Pi^{\tau}$ (\ref{eq:pitau}) with respect to the displacement field yields the weak form of the mechanical equilibrium equation
\begin{align} \label{eq:weaku}
 \delta_u \Pi^\tau & = \diffp{}{\ve{u}} \Pi^\tau \delta\ve{u} + \diffp{}{\nabla\ve{u}} \Pi^\tau \delta\nabla\ve{u} = \inte{\mathcal{B}}{\left[ \sig : \delta\eps - \bar{f}\cdot\delta \ve{u} \right]}{v} - \inte{\partial\mathcal{B}^\mathrm{N}}{\bar{\tau}\cdot \delta\ve{u}}{a} =0 
\end{align}
with the variations of displacement and strain $\delta \ve{u}$ and $\delta\eps$.
Applying \textsc{Gauß}' theorem retrieves its local form
\begin{equation}
	\nabla\cdot\sig + \ve{\bar{f}} = \ve{0} \text{ in } \calB
\end{equation}
with the boundary condition $ \sig\cdot\ve{n}=\ve{\bar{t}}$ on $\partial\calB^\mathrm{N}$ with $\partial\calB = \partial\calB^\mathrm{D} \cup \partial\calB^\mathrm{N}$ and $ \varnothing  = \partial\calB^\mathrm{D} \cap \partial\calB^\mathrm{N} $.

\subsubsection*{Plasticity}

Variation with respect to the plastic variables yields
\begin{align} \label{eq:plas}
	\delta_p \Pi^\tau & = \diffp{}{\qal} \Pi^\tau \delta\qal + \diffp{}{\dot{\te{q}}_{\alpha}} \Pi^\tau \delta\dot{\te{q}}_{\alpha} \\
	& = \inte{\calB}{\left\{\diffp{\psi}{\qal}\,\delta\qal + \intee{t_n}{t}{\left[\diffp{\phi^\mathrm{p}}{\qal}\delta\qal + \diffp{\phi^\mathrm{p}}{\dot{\te{q}}_{\alpha}}\delta\dot{\te{q}}_{\alpha}\right]}{\tau} \right\}}{v} \\
	& = \inte{\calB}{\left\{ \left( \diffp{\psi}{\qal} + \diffp{\phi^\mathrm{p}}{\dot{\te{q}}_{\alpha}} \right)\delta\qal + \intee{t_n}{t}{\left( \diffp{\phi^\mathrm{p}}{\qal} - \left(\diffp{\phi^\mathrm{p}}{\dot{\te{q}}_{\alpha}}\right)^. \right)\delta\qal}{\tau} \right\}}{v} =0.
\end{align}
Assuming the limiting case $t \rightarrow t_n$, the condition 
\begin{equation} \label{eq:biot}
	\diffp{\psi}{\qal} + \diffp{\phi^\mathrm{p}}{\dot{\te{q}}_{\alpha}} = 0
\end{equation}
must hold. This equation is known as \textsc{Biot}'s equation and is the basis for deriving the evolution of the plastic variables. For clarity, this is demonstrated with an exemplary dissipation potential taken from \textsc{Aygün} et al. \cite{aygun_coupling_2021a}
\begin{equation}
	\phi^\mathrm{p}(\dot{\epsp},\dot{\te{\alpha}}) = \sigma^\mathrm{y} ||\dot{\epsp}|| + \frac{b}{2} \left( \dot{\epsp} + \dot{\te{\alpha}} \right)^2, \quad \sigy =\mathrm{const.},\, \sigy > 0.
\end{equation}
The plastic set in this case contains $\qal=\{\epsp,\te{\alpha}\}$.  For the case $||\dot{\epsp}|| \neq 0$, it follows
\begin{align}
	\diffp{\psi}{\epsp} + \diffp{\phi^\mathrm{p}}{\dot{\epsp}} = 0: \quad & \sig = \diffp{\phi^\mathrm{p}}{\dot{\epsp}} = \sigy \frac{\dot{\epsp}}{||\dot{\epsp}||} + b\,(\dot{\epsp}+\dot{\te{\alpha}}) \label{eq:sig}\\
	\diffp{\psi}{\te{\alpha}} + \diffp{\phi^\mathrm{p}}{\dot{\te{\alpha}}} = 0: \quad & \te{\chi} = \diffp{\phi^\mathrm{p}}{\dot{\te{\alpha}}} = b\,(\dot{\epsp}+\dot{\te{\alpha}}). \label{eq:chi}
\end{align}
From the difference (\ref{eq:sig})$-$(\ref{eq:chi}) we get 
\begin{equation}
	\sig-\te{\chi} = \sigy \frac{\dot{\epsp}}{||\dot{\epsp}||}.
\end{equation}
Defining the plastic multiplier $\lambda = ||\dot{\epsp}||$ and the yield function $f^\mathrm{p} = ||\sig-\te{\chi}||-\sigy$ we obtain
the evolution equation for the plastic strain and the \textsc{Karush-Kuhn-Tucker} (KKT) conditions 
\begin{equation}
	\dot{\epsp} = \lambda \diffp{f^\mathrm{p}}{\sig} \quad \text{and} \quad \lambda\geq0, \, f^\mathrm{p}\leq0, \, \lambda f^\mathrm{p}=0.
\end{equation} 
Subsequently, the consistency condition $\lambda\dot{f}^\mathrm{p}=0$ follows from the KKT. For a detailed derivation including both cases $||\dot{\epsp}|| \neq 0$ and $||\dot{\epsp}||=0$ see \ref{ap:convexanalysis}. Please note that this model happens to be rate-dependent and was chosen only due to its simple structure. 
Further, see \ref{ap:maxdiss} for an alternative way of deriving the plastic model equations via a dissipation potential following the principle of maximum dissipation.



\subsubsection*{Phase-field}

Stationarity conditions w.\,r.\,t. the phase-field variable yield the weak form, here for the example AT 2,
\begin{align} \label{eq:weakd}
\delta_d \Pi^\tau =& \diffp{}{d} \Pi^\tau \delta d + \diffp{}{\nabla d} \Pi^\tau \delta\nabla d = 0 \\
= &\int_{\calB}\left\{\left[g'(d)\left(\psie_+ + \psip +\Delta^\mathrm{p}\right) +{\color{red}\tilde{g}'(d)H(\calF)}  +{\color{blue}h(\calF)}\frac{\Gc}{\ell}d  
\right] \delta d\, \right.  \nonumber\\
& \quad + {\color{blue}h(\calF)}\Gc\ell d_{,l} \, \delta d_{,l} + \intee{t_n}{t}{\eta\dot{d}\,\delta\dot{d}}{\tau}\biggr\}\mathrm{d}v, 
\end{align}
further demanding $\dot{d}\geq0$. The limiting case $t \rightarrow t_n$ now leads to the evolution equation
\begin{equation} \label{eq:ev}
\eta \dot{d} =  \Gc{\color{blue}h(\calF)}\left(\ell\Delta d - \frac{d}{\ell} \right) + {\color{blue}\Gc\ell\nabla d\nabla h(\calF)}   -g'(d){\left(\psie_+ + \psip +\Delta^\mathrm{p} \right)} {\color{red}-\tilde{g}'(d) H(\calF)} 
\end{equation}
and the boundary condition $\nabla d \cdot \ve{n} = 0$.
In order to ensure $\dot{d}\geq0$, most models use the history variable approach \cite{miehe_phase_2010}. Adopting the prevalent case $\tilde{g}(d)=g(d)$, the history variable $\calH$ can be introduced as
\begin{equation} \label{eq:evomax}
	\eta \dot{d} = \Gc{\color{blue}h(\calF)}\left(\ell\Delta d - \frac{d}{\ell}\right) {\color{blue} + \Gc\ell\nabla d\nabla h(\calF)} -g'(d) \underbrace{\max_{\tau\in[0,t]} \left(\psie_+(\tau) + \psip(\tau) {\color{red} + H(\calF,\tau)} + \Delta^\mathrm{p}(\tau)\right)}_{\mathcal{H}}  .
\end{equation}
This formulation is actually not variationally consistent. See appendix \ref{ap:penalty} for an alternative penalisation approach proposed by \cite{gerasimov_penalization_2019}.

\subsubsection*{Model equations and variables}

As shown, the variational principle yields a general set of governing equations. All model variables are displayed in Table \ref{tab:variables}, while Table \ref{tab:equations} lists all resulting model equations.

\begin{table}[h]
	\caption{Overview of model variables and their respective conjugate variables for general phase-field framework for fatigue fracture.} \label{tab:variables}
	\begin{tabular}{lll}
		 			& \textbf{Variable} & \textbf{Conjugate variable} \\ \midrule
		 Elasticity & Displacement $\ve{u}$ & \\
		  			& Elastic strain $\epse$ & $\displaystyle \sig = \diffp{\psi}{\epse}$ \\ \midrule
		 Plasticity & Plastic strain $\epsp$ & $\displaystyle  \sig = -\diffp{\psi}{\epsp}$ \\
		 			& Kinematic hardening variable $\te{\alpha}$ & $\displaystyle  \te{\chi} = -\diffp{\psi}{\ve{\alpha}} $ \\
		 			& Isotropic hardening variable $\alpha$ & $\displaystyle  p = -\diffp{\psi}{\alpha}$\\  \midrule
		 Fracture	& Phase-field $d$ & $\displaystyle \zeta^d = -\diffp{\psi}{d}$ \\
		 			& Phase-field gradient $\nabla d$ & \\ \midrule
		 Fatigue	& Fatigue damage $\calF$ &
	\end{tabular}	
\end{table}

\begin{table}[h] 
	\caption{Overview of model equations for general phase-field framework for fatigue fracture.} \label{tab:equations}
	\begin{tabular}{ll}
					\toprule
					& Free energy density $ \displaystyle \psi=g\, \psie_+ + \psie_- +  \psip$ \\
					& Strain definition $\displaystyle \eps=\epse+\epsp = \frac{1}{2}\left( \nabla \ve{u} + \nabla\ve{u}^\top \right)$ \\ 
					& Stress $\displaystyle \sig = \diffp{W_\mathrm{el}}{\epse}$ \\ \midrule
		 Equilibrium & Equilibrium $\displaystyle \nabla\cdot\sig + \ve{\bar{f}} = \ve{0}$\\
		 			& Boundary conditions $ \sig\cdot\ve{n}=\ve{\bar{t}}$ on $\partial\calB^\mathrm{N}$, $\ve{u}=\bar{\ve{u}}$ on $\partial\calB^\mathrm{D}$  \\ \midrule
		 Plasticity & Hardening variables $\displaystyle \te{\chi} = -\diffp{\psi}{\ve{\alpha}}, p = -\diffp{\psi}{\alpha}$\\
		 			& Yield function $\displaystyle f^\mathrm{p} (\sig,\te{\chi},p)$, often 
		 			$f^\mathrm{p}= \sqrt{\frac{3}{2}||\dev(\sig)-\dev({\te{\chi}})||^2} - \sigma^\mathrm{y} + p
		 			$\\
		 			& Flow rules and hardening laws, often $\dot{\epsp}=\lambda\ve{n}_\mathrm{p},\,\dot{\te{\alpha}}=\lambda\ve{n}_\mathrm{p},\,\dot{\alpha}=\lambda$ with $\ve{n}_\mathrm{p} = \frac{\frac{3}{2}\left( \dev\sig-\dev\te{\chi} \right)}{\sqrt{\frac{3}{2}||\dev\sig - \dev\te{\chi}||^2}}$
		 			\\
		 			& KKT, consistency condition $f^\mathrm{p}\leq0,  \lambda\geq0,  f^\mathrm{p}\lambda =0, \lambda\dot{f^\mathrm{p}}=0$\\ \midrule
		 Fracture 	& Evolution equation (including yield function)  \\
		 			& $\displaystyle \eta\dot{d} =  \Gc{\color{blue}h(\calF)}\left(\ell\Delta d - \frac{d}{\ell} \right) + {\color{blue}\Gc\ell\nabla d\nabla h(\calF)}   + 2(1-d){\left(\psie_+ + \psip  \right)} {\color{red}-\tilde{g}'(d) H(\calF)} - g'(d)\Delta^\mathrm{p} = f^d$ \\
		 			& + irreversibility $\dot{d}\geq0$ \\
		 			& \underline{or} KKT $f^d\leq0, \dot{d}\geq0, f^d \dot{d} =0$ \\
		 			& Boundary conditions $\nabla d \cdot \ve{n} = 0$\\ \midrule
		 Fatigue 	& Evolution of fatigue variable $\dot{\calF}$
	\end{tabular}	
\end{table}

\subsection{Alternative ways of model derivation}

Apart from the incremental variational principle presented here, there are many other ways to derive a phase-field fracture model and the publications mentioned in this paper already cover a wide variety of derivation methods. Since this can impede the comparison of models, it is helpful to demonstrate the analogies and how the different approaches are intertwined. Figure \ref{fig:scheme} gives an overview of different paths for model derivation for a phase-field model for fatigue fracture, possibly also including elastic-plastic material behaviour. The derivation used in Section \ref{sec:framework} is highlighted in red ("Way 1"). Quantities that can serve as starting points for general modelling choices are marked in blue. Although it is beyond the scope of this paper to repeat the derivation of the model with all strategies, a few common approaches are listed in the following:

\begin{itemize}
	\item The plastic dissipation potential can be derived by first setting the yield condition and using it then as a constraint for the optimisation following the principle of maximum dissipation. See \ref{ap:maxdiss}. Marked in green as "Way 2" in Figure \ref{fig:scheme}.
	\item Not only elastic-plastic material behaviour, but also the phase-field problem can be modelled using yield equations. \textsc{Noii} et al. \cite{noii_bayesian_2021a} show that the evolution equation of the phase-field model can be reformulated to $\eta\dot{d}=f^d$. The yield function $f^d$ being the difference between a (crack) driving force and a (crack) resisting force offers convenient starting point for modelling decisions due to its physical interpretability. See also \textsc{Miehe} et al. \cite{miehe_phasefield_2016} for a formulation based on yield functions for both phase-field and plasticity.
	\item The energetic formulation based on a local stability condition and a local energy balance is also a popular way to derive the set of model equations, as shown in \cite{alessi_phenomenological_2018}. See Figure \ref{fig:scheme}, top right corner. 
\end{itemize}

\newgeometry{margin=1cm}
\begin{landscape}
	\thispagestyle{empty} 
	
	\begin{figure} [h] 
		\includegraphics[width=\linewidth]{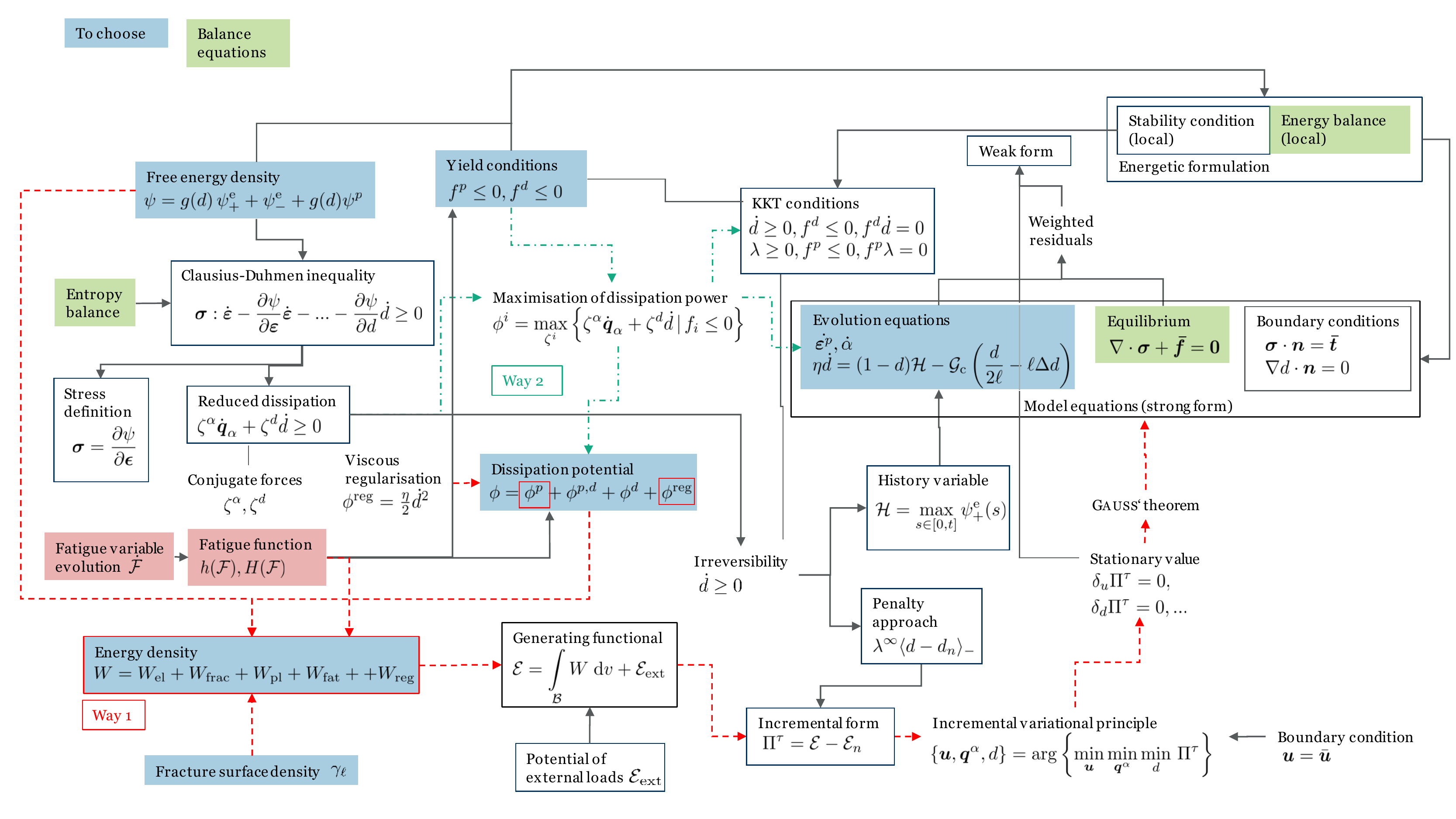}
		\caption{ Scheme of different ways of model derivation for a phase-field model for fatigue fracture, explicitly covering elastic-plastic material behaviour.  Balance equations are marked in green, while quantities suitable for implementing general modelling choices are marked in blue. Derived quantities are white. Highlighted are two possible ways of deriving model equations. }
		\label{fig:scheme}
	\end{figure}
		
\end{landscape}
\restoregeometry

\clearpage
\includepdf[pages=1,fitpaper]{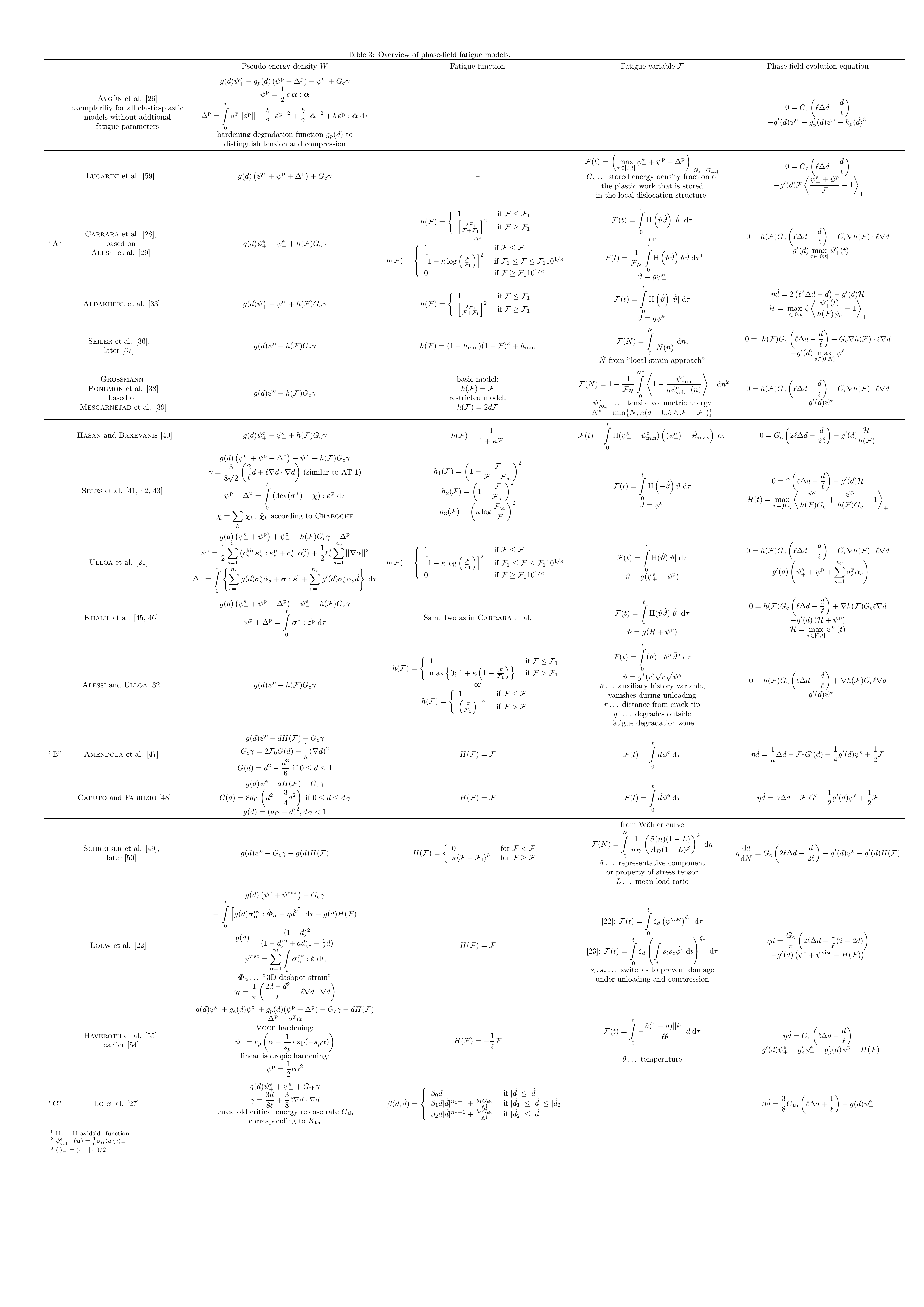}

\section{Overview of models}
\label{sec:overview}

This section gives an overview of most phase-field models for fatigue fracture published to date. If models share a very similar structure, only one of them is chosen as the representative example. Energy density $W$, fatigue function $h(\calF)$ or $H(\calF)$, fatigue variable $\calF$ and phase-field evolution equation of each model, according to the unified notation introduced in Section \ref{sec:framework},  are listed in Table 3. Please note that a similar table for $A$-models is presented in \cite{alessi_endowing_2023}. Additionally, a short description of each is given in the following. The models are categorised into {\color{blue}type \textit{A}} and {\color{red}type \textit{B}} according to their distinct fatigue terms as introduced in Section \ref{sec:framework}, and those that have a unique structure that does not belong to either of the aforementioned categories. 

\subsection{Type A}
\label{sec:list}

\subsubsection*{\textsc{Carrara} et al. \cite{carrara_framework_2019}}

This model was one of the first $A$-type models to be published. It is essentially a generalisation of the model by \textsc{Alessi} et al. \cite{alessi_phenomenological_2018} to 3D. It is a purely elastic model and therefore suitable for brittle material behaviour and HCF.
Due to its general and simple nature, many of the following $A$-type models refer to this one. Both the fatigue degradation functions and the fatigue variable based on the accumulated strain energy density have been  used in other models. The fatigue variable accumulates only during loading which is ensured by a Heaviside function. $\calF$ starts accumulating from the first load cycle. This has to be considered during model fitting in order to be consistent also during static loading. The authors were able to show that applied mean load can shift the \textsc{Paris} curve and that parameter $\kappa$ in the fatigue degradation function controls the \textsc{Paris} parameters $C$ and $m$.

\subsubsection*{\textsc{Aldakheel} et al. \cite{aldakheel_phasefield_2022}} 

This model is very similar to \textsc{Carrara}'s. Another model with the same structure but for piezo-electric materials and therefore a coupling to an electric field was published in \textsc{Tan} et al. \cite{tan_phase_2022}. {\textsc{Yin} and \textsc{Kaliske} \cite{yin_fatigue_2023} use the same structure and fatigue variable with slightly different fatigue degradation function for an elastomer material with \textsc{Neo-Hooke}ian elastic material behaviour.}

\subsubsection*{\textsc{Seiler} et al. \cite{seiler_efficient_2020,seiler_phasefield_2021a}}

The fatigue variable of this model is formulated in cycle domain rather than in time domain, describing the fatigue process continuously instead of simulating each loading and unloading phase. Therefore, a representative (often constant) loading instead of an oscillatory loading is used, saving computational time. See Section \ref{sec:cycles} for further explanation. Fatigue damage is calculated based on a structural durability concept which requires \textsc{Wöhler} curves as an input. The local elastic-plastic stress-strain state is approximated with the help of cyclic stress-strain curves. This allows for a simplified modelling of crack tip plasticity as long as the plastic zone stays small. The model is therefore especially suitable for the transitional range between LCF and HCF.

\subsubsection*{\textsc{Grossmann-Ponemon} et al. \cite{grossman-ponemon_phasefield_2022}}

This model is based on \textsc{Mesgarnejad} et al. \cite{mesgarnejad_phasefield_2019}. Its fatigue variable is also of a continuous type, formulated in cycle domain. The model parameters depend on the load ratio $R$ between load minimum and maximum within a load cycle, which has to be specified as an input. Therefore, the model reproduces mean load effects. It is evaluated both for a cubic and the standard AT 2 degradation function $g(d)$. Fatigue accumulation is inhibited in strongly degraded areas with $d\geq0.5$, thereby preventing further degradation within the zone of very high strain energy density. In another model variant, fatigue accumulation is only allowed in non-intact material where $d>0$.
The earlier paper of \textsc{Mesgarnejad} et al. \cite{mesgarnejad_phasefield_2019} proposed a formulation which degraded only the $d$- and not the $\nabla d$-term of the crack surface density. 

\subsubsection*{\textsc{Hasan} and \textsc{Baxevanis} \cite{hasan_phasefield_2021}}

Although formulated with a fatigue degradation function $h(\calF)$, this model resembles the structure of a $B$-type model. This becomes obvious when stating the evolution equation with $h(\calF) = 1/(1+\kappa\calF)$ 
\begin{equation}
	0 = h(\calF) \Gc \left( 2\ell\Delta d - \frac{d}{2\ell} \right) - g'(d)\calH = \Gc \left( 2\ell\Delta d - \frac{d}{2\ell} \right) - g'(d) (1+\kappa\calF) \calH
\end{equation}
if the gradient term $\nabla h(\calF)$ is neglected. The model is consistent for monotonic loading (see also Section \ref{sec:fatvar}) and is able to reproduce both \textsc{Paris} and \textsc{Wöhler} behaviour.

\subsubsection*{\textsc{Sele\v{s}} et al. \cite{seles_numerical_2020,seles_general_2021,seles_microcrack_2021}}

This model includes an elastic-plastic material law with isotropic and kinematic hardening of \textsc{Chaboche} type. Since the accumulating plastic strain energy density $\psip$ is part of the crack driving force, plastic processes promote crack growth regardless of the fatigue variable, which only depends on the elastic strain energy density $\psie_+$. In this way, this model covers both LCF -- driven by plastic strains -- and HCF due to small stress amplitudes which cause no macroscopic plastic effects. The model automatically reproduces mean load effects due to the nature of its fatigue variable and \textsc{Paris} behaviour with its parameter $\calF_{\infty}$ controlling the \textsc{Paris} parameter $C$. Consistency with monotonic loading is ensured by accumulating fatigue damage only during unloading. In order to reduce computational time, a cycle skipping technique by \textsc{Cojacaru} and \textsc{Karlsson} \cite{cojocaru_simple_2006} is applied, see also Section \ref{sec:cycles}. This is of particular importance for ductile phase-field models which have even higher computational times than brittle ones. 

\subsubsection*{\textsc{Ulloa} et al. \cite{ulloa_phasefield_2021}}

Here, the model formulation is also based on the framework by \textsc{Alessi} \cite{alessi_phenomenological_2018}. This ductile phase-field model includes multi-surface kinematic hardening, gradient-enhanced isotropic hardening and softening as well as an explicit ratchetting strain variable. In contrast to \textsc{Sele\v{s}}, the fatigue variable accumulates from both the elastic and plastic strain energy density $\psie_+$ and $\psip$, strengthening the influence of plastic strains on the crack evolution. Again, LCF is mainly driven by plastic strains while HCF is driven by the fatigue variable. The first load cycle already leads to fatigue degradation.

\subsubsection*{\textsc{Khalil} et al. \cite{khalil_phasefield_2021,khalil_generalised_2022}}

This model is an extension of \textsc{Carrara}'s model to elastic-plastic material behaviour described by a \textsc{Chaboche} model with isotropic and nonlinear kinematic hardening. The fatigue variable accumulates from the the current temporal maximum $\calH=\max_t\psie_+(t)$ and the plastic strain energy density $\psip$. The general model formulation can recover both AT 1 and AT 2 as well as a cohesive zone model for suitable parameter choices. Instead of the staggered solution scheme used most frequently, the authors present a new pseudo-monolithic quasi-\textsc{Newton} scheme.

\subsubsection*{\textsc{Alessi} and \textsc{Ulloa}  \cite{alessi_endowing_2023}}

The authors introduce a new class of phase-field fatigue models with a strong link to fracture mechanics. Due to elastic material behaviour, they are suitable for HCF only. Still, microstructural ductile effects around the crack tip are acknowledged by introducing a fatigue degradation zone. Following the idea that for HCF these effects are limited to a small zone around the crack tip, fatigue damage is only accumulated within the zone, covering microstructural effects in a phenomenological way. The authors formulate four requirements to the model's behaviour which are met by four functions contributing to the fatigue variable. See Section \ref{sec:fatvar} for details. Using the example of a stationary crack they are able to correlate analytical and numerical results from their model with the \textsc{Paris} law. Thereby, they establish direct relations between model parameters and model behaviour, e.\,g. mean stress dependence and incline of the \textsc{Paris} curve can be controlled by a parameter each.

Relying on \textsc{Griffith}s fracture theory, they are able to establish a new solution strategy: In each increment in which the energy release rate $G$ satisfies $G\leq h(\calF)\Gc$, no crack propagation can take place and only $\calF$ is accumulated. If instead $G> h(\calF)\Gc$, the solution is not admissible. In that case, a solution is seeked under the condition $G= h(\calF)\Gc$.
The fatigue crack growth shows three stages: Initial damage accumulation, transient evolution of the crack and, finally, stable crack propagation. 

%
%
%
%

\subsection{Type B}

\subsubsection*{\textsc{Amendola} et al. \cite{amendola_thermomechanics_2016}}

This model is derived in \textsc{Ginzburg-Landau} form. The additive contribution to the crack driving force is controlled by a fatigue variable depending on the strain energy density. The authors also present a model variant for the non-isothermal case.

\subsubsection*{\textsc{Caputo} and \textsc{Fabrizio} \cite{caputo_damage_2015}}

This model is very similar to \textsc{Amendola}'s apart from the stress definition and the fracture surface density.

\subsubsection*{\textsc{Schreiber} et al. \cite{schreiber_phase_2020b,schreiber_phase_2021a}}

This $B$-type model obtains its fatigue damage from \textsc{Wöhler} curves. The application of representative loads instead of cycle-wise simulations allows for an accelerated computation. An efficient control for the number of load cycles per increment is presented. The effect of mean loads can be incorporated by using the mean load ratio of the external load in damage accumulation. Being a brittle model, it is only suitable for HCF. 

Since the fracture and fatigue contributions are interpreted as part of the free energy density, additional stress terms arise from the second law of thermodynamics.
With the fatigue variable $\calF(\eps)$ depending on the strain due to the empirical fatigue concept used, the stress has to be defined as
\begin{equation}
	\sig = g(d) \mathbb{C} \eps + g(d)qb\langle \calF-\calF_{\min}\rangle^{b-1} \diffp{\calF}{\eps}.
\end{equation}
The additional stress contributions are interpreted as micro stresses due to microscopic fatigue mechanisms. See Section \ref{sec:stress} for more details. This model was extended to incorporate thermal effects in \textsc{Yan} et al. \cite{yan_simulating_2022} { and derived in the framework of configurational forces in \textsc{Yan} et al. \cite{yan_configurational_2023}, showing that the fatigue contribution yields an additional configurational force itself}.

\subsubsection*{\textsc{Loew} et al. \cite{loew_fatigue_2020}}

This model is -- in contrast to most other models described here -- meant not for metals but for rubber. Due to the nature of this material, the model is formulated in a large strain setting and is of viscous, i.\,e. rate-dependent nature. The stress 
\begin{equation}
	\sig  = g(d) \left( \sig^\mathrm{eq} + \sum_{\alpha=1}^{m} \sig^\mathrm{ov}_\alpha \right)
\end{equation}
contains therefore an additional overstress part. Although this model is a $B$-type model, in an earlier publication \cite{loew_fatigue_2019}, the authors introduced a model variant without an additional fatigue term, where fatigue fracture was exclusively driven by viscous effects, in the form of an accumulating viscous energy density. See the model \textsc{Aygün} et al. \cite{aygun_coupling_2021a} below for a similar concept in plasticity. However, the newer publications include the viscous strain energy density $\psi^\mathrm{visc}$ not only in the crack driving force but also in an additional fatigue variable, strengthening the effect of viscosity on fatigue crack growth. A cycle jump technique is used (at least for the elastic case), introducing an explicit and an implicit acceleration scheme with adaptive jump control \cite{loew_accelerating_2020}.

\subsubsection*{\textsc{Haveroth} et al. \cite{boldrini_nonisothermal_2016,haveroth_nonisothermal_2020}}

The authors present a comprehensive model framework including plasticity covering non-isothermal conditions and time-rate and inertia effects. Fatigue is incorporated as an extra (phase) field variable with its evolution equation derived from the second law of thermodynamics instead of applying a phenomenological evolution law like most other models. The fatigue phase-field is interpreted as micro-damage variable covering micro-cracks and -voids while the regular phase-field for fracture describes macro- and meso-cracks. 
Simulations can be accelerated through a cycle jump technique. 

\subsection{Other models}

\subsubsection*{\textsc{Aygün} et al. \cite{aygun_coupling_2021a}}

This model is included exemplarily for all standard ductile phase-field models which model fatigue effects without an explicit fatigue variable. Fracture is only driven by an accumulating plastic energy density $\psip$ in the crack driving force. In this case, an \textsc{Armstrong-Frederick} elastic-plastic material law is used. Naturally, these types of models are only suitable for LCF with significant plastic strains. This model is rate-dependent. 
Another example for a fatigue model without a fatigue variable is \textsc{Schröder} et al. {\color{blue}\cite{schroder_phasefield_2022,pise_phenomenological_2023}} for concrete or cementious materials. \textsc{Tsakmakiks} and \textsc{Vormwald} \cite{tsakmakis_phase_2023} also showed the ability of their ductile phase-field model derived in the framework of so-called non-conventional thermodynamics to cover fatigue fracture.

{
\subsubsection*{\textsc{Lucarini} et al. \cite{lucarini_fftbased_2023}}

In a first attempt to bridge the scales, this model is designed for fatigue fracture at the microscale. Applied to a representative volume element (RVE) of a microstructure, this model supposes periodic fields. Therefore, a FFT-based solver can be applied to increase the performance. The crack driving force $\calH$ is reformulated completely. It now depends on the stored energy density according to a crystal plasticity model by \textsc{Dunne} et al. \cite{dunne_lengthscaledependent_2007}, being the fraction of the plastic work that is stored in the local dislocation structure. The model derives the plastic shear strain rate from a mechanistic slip rule and the critical resolved shear stress follows a hardening law. They were able to show fatigue fracture in singular crystals, following preferred slip planes, and multi-crystal simulations with transgranular crack growth.



}

\subsubsection*{\textsc{Lo} et al. \cite{lo_phasefield_2019}}

This model represents a totally different type of phase-field fatigue model. Unlike the $A$- and $B$-type models, neither additional crack driving terms nor degradation of the fracture toughness are used. Instead, the viscosity parameter $\eta$, which is only seen as a numerical damping parameter in most other quasi-static phase-field models, controls the fatigue crack growth. It is a function fitted to \textsc{Paris} curves which serve as an input for the model. No cycle-wise simulation is performed, the load is applied statically instead. The model uses a linear approximation of the crack surface.

\section{Discussion} 

This section discusses the most important characteristics and modelling choices, which present similarities and differences between the models listed in the previous section. Apart from an example for the differentiation in $A$- and $B$-models, the comparison remains on a theoretical level. For extensive numerical examples we refer to the original publications.

\subsection{A- and B-models} 

The most distinct feature of the models is the way their fatigue variable is implemented in an originally static phase-field model. Most models reviewed here are of either the $A$- or $B$-type as introduced in Section \ref{sec:framework}. In the following, a numerical comparison between the two is performed in order to investigate the behaviour of both model types in a cyclic simulation.

\subsubsection{Numerical setup} 

\begin{figure}[t]
	\begin{minipage}[c]{0.74\textwidth}
		\renewcommand{\arraystretch}{2} 
		\begin{tabular}{l|c|c}
			\hline
			&	\textbf{Model $A$} & \textbf{Model $B$} \\ \hline
			\makecell{\textbf{Elastic} \\parameters} & \multicolumn{2}{c}{$E=210$ GPa, $\nu=0.3$ } \\ \hline
			\makecell{\textbf{Fracture} \\parameters} & \multicolumn{2}{c}{ \makecell{$\Gc=0.039$ kN\,mm$^{-1}$, $\ell=1$ mm, $\eta=10^{-7}$ GPa\,s,\\ no split of elastic strain energy}} \\ \hline
			\makecell{\textbf{Fatigue} \\variable} &  \multicolumn{2}{c}{\makecell{From local strain approach as in \textsc{Seiler} et al. \cite{seiler_efficient_2019} \\ with $\sigma'_f = 735$ MPa, $\varepsilon'_f=0.59$, $B=-0.087$, $c=-0.58$, \\$K'=796$ MPa, $n=0.15$}} \\
			\hline
			\makecell{\textbf{Fatigue} \\ {function}} & \makecell{$h(\calF) = (1-h_{\min})/(1-\calF)^\kappa$\\$+h_{\min}$ \\ $h_{\min}= 0.05$, $\kappa=1$} & \makecell{ $W^B_\mathrm{fat} = g(d) H(\calF) = g(d) b \calF^{\xi}$ \\ $b=2\cdot10^{-2}$ GPa, $\xi=20$ }               \\ \hline
		\end{tabular}
	\end{minipage}
	\begin{minipage}[c]{0.25\textwidth}
		\def\svgwidth{\linewidth}\small{
			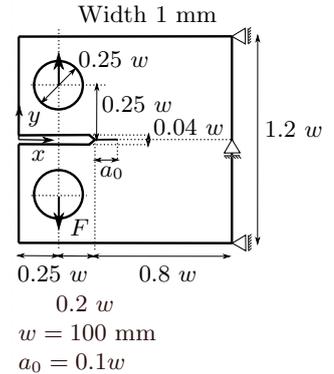}
	\end{minipage}
	\caption{Material parameters (left) and geometry of CT specimen (right) for simulations with models $A$ and $B$.
	}\label{fig:numex}
\end{figure}

Both models are tested with a Compact Tension (CT) geometry displayed in Figure \ref{fig:numex} with assumend plane strain state. The initial crack is applied as a \textsc{Dirichlet} condition for the phase-field. The mesh is refined in the area of crack growth to a minimum element size of 0.3 mm. The specimen is loaded with load cycles of constant force amplitude with maximum load $F=2$\,kN and a load ratio between minimum and maximum load $R=-1$. Construction steel is assumed as a material, {the material model is purely elastic. The corresponding elastic and fracture parameters are listed in Figure \ref{fig:numex}. $\Gc$ is usually determined from CT tests, although some compute it from the maximum tensile strength in 1D \cite{kuhn_degradation_2015}. Regarding the characteristic length $\ell$, two perspectives exist: It is either seen as a numerical parameter and therefore chosen as small as possible, or as a material parameter characterising the sharp or more diffuse crack tip, e.\,g. due to pores. The recommendations for static phase-field simulations regarding the ratio between $\ell$ and the element size cannot be applied to $A$- and $B$-type models, since the regularisation profile is disturbed, as will be explained later on. }

As a fatigue variable, the one by \textsc{Seiler} et al. \cite{seiler_efficient_2019} based on the local strain approach is chosen exemplarily for both models. See also Figure \ref{fig:numex} for the parameters used to determine the fatigue variable, {which are taken from cyclic tensile experiments}. The fatigue degradation function for model $A$ is set as in \cite{seiler_efficient_2019} to
\begin{equation}
	h(\calF) = (1-h_{\min})/(1-\calF)^\kappa+h_{\min}
\end{equation} 
while the additive energy term for model $B$ is
\begin{equation}
	W^B_\mathrm{fat} = g(d)\, H(\calF) = g(d)\, b\, \calF^{\xi},
\end{equation}
see Figure \ref{fig:numex} for parameters. {These parameters have to be calibrated to experiments.} Fatigue variable and fatigue functions are chosen arbitrarily and are not the subject of this analysis of the model types. The coupled problem is solved using an alternate minimisation algorithm with error control for the iteration over both fields.

\subsubsection{Results and discussion} 

\begin{figure} [h] 
	\def\svgwidth{\linewidth}\small{
		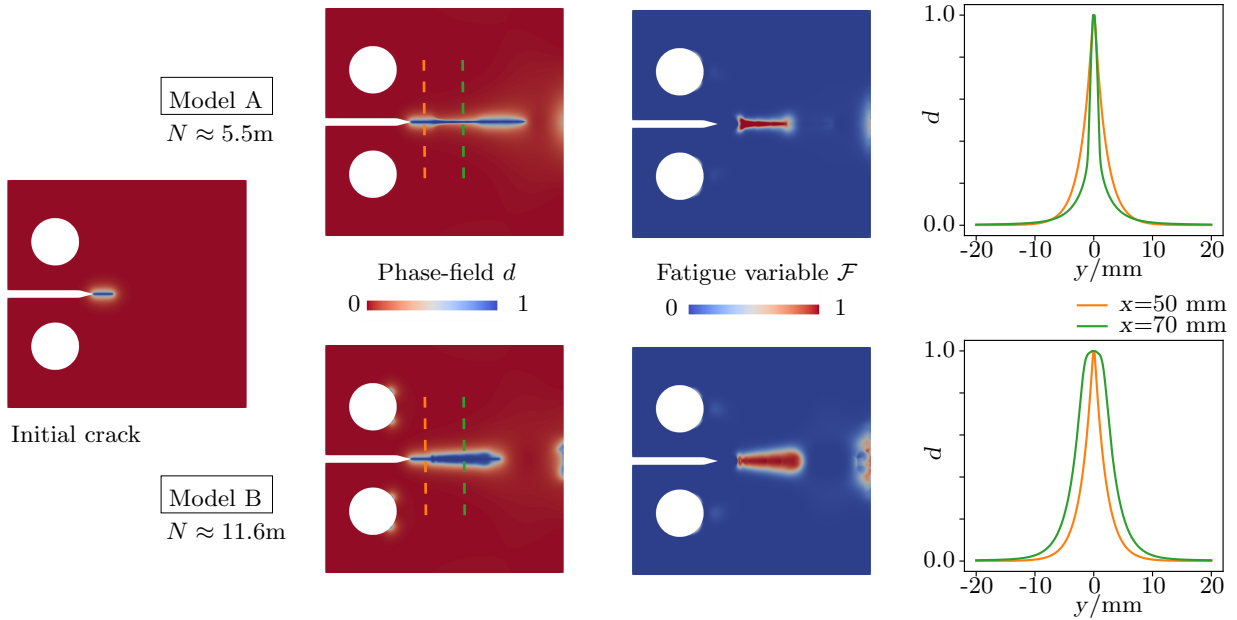}
	\caption{ Results of phase-field fatigue simulation with $A$- and $B$-model. Initial setup with pre-defined phase-field crack and results after simulation of $N$ load cycles. Cross sections on the right show phase-field profile within ideal pre-defined crack (orange) and fatigue crack (green). Model $A$ narrows the phase-field profile compared to static crack while model $B$ leads to widening of the phase-field crack. }
	\label{fig:results}
\end{figure}

Figure \ref{fig:results} shows the simulation results for both models after different amounts of load cycles. The difference in total lifetime is a matter of parametrisation and not due to the model types. The pre-existing initial crack grows into a fatigue crack with stable cyclic crack growth, until, finally, it evolves into unstable residual fracture. In this last stage, the crack proceeds under monotonic load without evolution of the fatigue variable, as becomes apparent from the distribution of the fatigue variable. Both the initial crack and the residual crack show an ideal regularised phase-field profile determined by the characteristic length scale $\ell$. \textsc{Miehe} et al. \cite{miehe_thermodynamically_2010} demonstrated analytically that this regularisation has to be of exponential nature in order to be a solution to the phase-field differential equation. The profile of the initial crack phase-field is plotted in the diagram on the right of Figure \ref{fig:results}, marked in orange. 

However, the section of cyclic crack growth shows different profiles for the two model versions. Model $A$ yields a -- compared to the ideal crack -- narrowed crack profile. This becomes evident in both the phase-field contour plot and the green graph in the diagram on the right, Figure \ref{fig:results}. It can be explained with the weak form of the phase-field problem (here for the elastic case)
\begin{align} \label{eq:modelA}
	0 = &\int_{\calB}\left\{\left[g'(d) \psie_+  +{\color{blue}h(\calF)}\frac{\Gc}{\ell}d  
	\right] \delta d\,    + \underline{ {\color{blue}h(\calF)}\Gc\ell\, \nabla d \, \delta (\nabla d) } + \intee{t_n}{t}{\eta\dot{d}\,\delta\dot{d}}{\tau}\right\}\mathrm{d}v.
\end{align}
The fatigue degradation function $h(\calF)$ reaches very low values $\ll1$ in most parametrisations in the literature, here its minimal value is $h_{\min}=0.05$. It affects the phase-field gradient term underlined in the equation above. This term is meant to regularise the problem and thereby controls the shape of the phase-field profile. When this term is now weakened due to the fatigue degradation function, the profile develops more freely. In the present case this leads to a narrowing of the crack profile as the crack evolves within the narrow "corridor" of lowered fracture toughness controlled by the fatigue variable $\calF$, see also its contour plot.

\textsc{Grossmann-Ponemon} et al. \cite{grossman-ponemon_phasefield_2022} and \textsc{Hasan} and \textsc{Baxevanis} \cite{hasan_phasefield_2021} also observe this crack narrowing compared to the brittle model. Irregularities due to heterogeneous or non-constant $\Gc$ appear not only in fatigue models. See e.\,g. \cite{dammass_phasefield_2022a} for a rate-dependent fracture toughness and \cite{hansen-dorr_phasefield_2018} for the effect an inhomogeneous distribution of $\Gc$ has on the effective crack resistance, as well as \cite{alessi_endowing_2023} for a discussion on how they consider $\Gc$ not as a material parameter but a material function and an overview of different reasons for non-constant $\Gc$. 

Model $B$, on the other hand, shows a widening of the crack profile. In the phase-field evolution equation of this model variant
\begin{equation} \label{eq:modelB}
	\eta \dot{d} = \Gc\left(\ell\Delta d - \frac{d}{\ell}\right)  -g'(d) \underbrace{\left(\psie_+ {\color{red} + H(\calF)}\right) }_{\mathcal{H}}
\end{equation}
the fatigue term appears within the crack driving force $\calH$. This leads to a very direct coupling between the fatigue contribution $H(\calF)$  and the phase-field distribution. The contour plots of phase-field $d$ and fatigue variable $\calF$ therefore show a very similar distribution. Hence, due to the strong coupling, the nature of the fatigue variable is even more decisive for the crack appearance for the $B$-type model than it is for other model classes. Then again, the model type and fatigue function $h(\calF)$ and $H(\calF)$ also influence the fatigue variable, which becomes clear from the two different distributions of $\calF$ -- which is in this case derived from the strain -- for the two model versions. \textsc{Schreiber} \cite{schreiber_phase_2021b} also observe the crack widening for their $B$-type model for small deviations of their ideal fatigue function.

Both the widening and the narrowing of the phase-field profile can lead to deviation of the crack energy which is not (and doesn't necessarily have to be) in accordance with the regularisation of static phase-field models. The crack growth rate can also be affected and responds sensitively to the nature and distribution of the fatigue variable and the fatigue function. {The degradation of the gradient term for the $A$-type model can lead to a mesh sensitivity. Also due to the different parameters influencing mesh sensitivity, e.\,g. threshold values of the fatigue degradation function, no general recommendation can be given regarding the mesh size, also not for the $B$-type model. Instead, a convergence analysis should be performed.}

An important difference between the model types is also that for $A$-type models, the fatigue degradation function $h(\calF)$ has obviously to be within the range $[0,1]$, whereas the $B$-type fatigue function $H(\calF)$ has no upper boundary and its order of magnitude must be calibrated during parametrisation. {Regarding computational time, the models do not differ significantly, for the two main time-consuming routines need the same amount of time: The assembly including the computation of the fatigue variable and the solution of the system of equations with the same number of degrees of freedom.}

As shown, both model types entail numerical difficulties reflected in their phase-field profile. The choice of a model variant should eventually be based upon the desired physical interpretation: Some model approaches and applications are suited for a reduction of the material's crack resistance while others go with an increase of the crack driving force compared to the static case.

\subsection{Fatigue variable} 
\label{sec:fatvar}

Besides the basic model structure, the fatigue variable $\calF$ is the second most important choice in the model. Most models studied here use either a variation of the accumulated strain energy density (\textsc{Carrara} et al. \cite{carrara_framework_2019}, \textsc{Grossmann-Ponemon} et al. \cite{grossman-ponemon_phasefield_2022}, \textsc{Loew} et al. \cite{loew_fatigue_2020} etc.) or an empirical fatigue concept  (\textsc{Schreiber} et al. \cite{schreiber_phase_2020b}, \textsc{Seiler} et al. \cite{seiler_efficient_2020}).

The energy density is an obvious choice due to its easy accessibility in a material routine. \textsc{Xu} et al. \cite{xu_microstructural_2021} explain its suitability from a microscopic point of view: The crack growth rate of short cracks depends on the microstructural crack path and the local crack propagation rate. Conveniently, the stored energy density happens to be a microstructure-sensitive driving force due to being a measure of the energy stored in the lattice structure available to eventually create new crack surface \cite{xu_microstructural_2021}. With a single crystal plasticity slip system, they show that the stored energy density depends on the Burgers vector and the critical resolved shear stress, two characteristics for the microstructure of the material. Furthermore, it is consistent with fracture mechanics, being related to the stress intensity factor which is shown to control fatigue crack growth \cite{xu_microstructural_2021}. 
They were also able to show experimentally that stored energy at the crack tip (determined with the help of DIC measurements) leads to a higher crack propagation rate. 

The models that use the strain energy density for the fatigue variable differ from each other regarding the conditions for damage accumulation. Some only accumulate during loading (when the micro cracks evolve, supposedly, \textsc{Carrara} et al. \cite{carrara_framework_2019}) or only during unloading (\textsc{Sele\u{s}} et al. \cite{seles_numerical_2020}) in order to be consistent with models for static loading: Loaded with a purely monotonic load, no fatigue damage should be accumulated. Moreover, most models use the degraded tensile strain energy density $g(d)\psie_+$. Some use it without degradation (\textsc{Sele\u{s}} et al. \cite{seles_numerical_2020}). In this case, $\calF$ accumulates further even when a phase-field crack has already formed.

\textsc{Alessi} and \textsc{Ulloa} \cite{alessi_endowing_2023} present a modular scheme to construct a fatigue variable based on the strain energy density in order to fulfill their four requirements towards the model behaviour. They are met by four functions contributing to $\calF$, respectively. Firstly, they treat the singularity of $\psie$ at the crack tip by smoothing it out within a certain zone, the fatigue degradation zone. Outside the zone, no (or close to no) fatigue variable is accumulated. This is meant to phenomenologically replicate the microstructural ductile effects, which mainly occur around the crack tip. Further, two functions specify the damaging loading types and the mean stress effect shifting the \textsc{Paris} curve in vertical direction, respectively. An additional exponential function controls the incline of the \textsc{Paris} curve. In this way, the phenomena of fatigue crack growth can be tuned individually.

The other group of models obtain their fatigue variable through empirical lifetime estimation concepts for engineering components. They use data from standardized experiments, i.\,e. \textsc{Paris} curves, \textsc{Wöhler} curves and strain \textsc{Wöhler} curves as input data. Conveniently, this incorporates additional information about the fatigue behaviour of the material into the model. However, the models still include parameters to be be fitted to experimental results, usually as a part of the fatigue function $h(\calF)$ or $H(\calF)$. Due to their underlying assumptions, these concepts allow for an accelerated model implementation, see Section \ref{sec:cycles}. 

In brief, the latter models use a damage evaluation based on remaining lifetime \cite{lemaitre_mechanics_1998}. This requires some sort of normalization of a lifetime describing variable. The former models based on the strain energy density, on the other hand, do without such a normalization and accumulate $\calF$ "en passant", but have to use more arbitrary parameters without a direct relation to experimental quantities. 

Multiaxial, possibly even non-proportional loads might call for fatigue variables which can replicate stressing and damage history varying in direction. Even though the strain energy density contains the full stress state, the expression lacks information of direction. Traditional life estimation concepts, on the other hand, are often applied with critical plane concepts \cite{karolczuk_review_2005}, accumulating fatigue damage for several discrete directions individually. So far, this has not been exploited yet for phase-field fatigue models, though.

\subsection{Fatigue functions $h(\calF)$ and $H(\calF)$}

The fatigue functions $h(\calF)$ and $H(\calF)$ usually contain the most important parameters for model fitting. Those are thresholds or control the progressive or degressive evolution of the fatigue contribution. In this way, they often influence the inclination and shift of the resulting \textsc{Paris} curve and/or \textsc{Wöhler} curve. The distinction between $A$- and $B$-models and therefore $h(\calF)$ and $H(\calF)$ functions is only for illustrative purposes: The two formulations can be converted into each other by a suitable choice of the functions. \textsc{Hasan} and \textsc{Baxevanis} \cite{hasan_phasefield_2021} chose $h(\calF)$ in a way that that creates a $B$-type model, see Section \ref{sec:list}. The other way around, by setting $H(\calF)=h(\calF)\Gc\gamma(d,\nabla d)$ one recovers the $A$-type model. $A$-type models describe the weakening of the material through a gradual decrease of fracture toughness $\Gc$. To date, all functions $h(\calF)$ are arbitrary choices since no model is based on an experimental measurement of the degrading fracture toughness yet.

\subsection{Treatment of plasticity}

Strictly speaking, the range of application of an elastic phase-field model is limited to brittle materials or ductile materials (such as metals) only for HCF. Therefore, plasticity is often incorporated in the models in the form of a plastic energy density $\psip$, describing the accumulated energy due to hardening. In the phase-field evolution equation (\ref{eq:ev}) it appears in the static crack driving force. Already this effect alone can describe cyclic material degradation under (comparatively high) cyclic loads leading to phase-field cracks, as shown in \textsc{Aygün} et al. \cite{aygun_coupling_2021a}. If combined with a fatigue variable depending on the elastic energy density, this can cover a wide range of loads from LCF to HCF. Some models double the effect of plasticity on the crack evolution by including $\psip$ also in the fatigue variable $\calF$ (\textsc{Ulloa} et al. \cite{ulloa_phasefield_2021}). This allows for more modelling flexibility and is motivated by the fact that plastic processes also drive static cracks (therefore ensuring consistency with monotonic loading) while at the same time, they influence fatigue qualities of the material, especially on the microscopic scale. Microscopic plastic effects have not been modelled explicitly so far since multiscale phase-field modelling of fracture remains a challenging task, i.\,a. due to being very computationally intensive.

\subsection{Irreversibility}

The problem of crack irreversibility is a frequently discussed matter in the phase-field community. Different approaches to ensure $\dot{d}>0$ (the strictest formulation) exist, such as the history variable approach and the penalty parameter, see Section \ref{sec:framework}. While most phase-field fatigue models use a history parameter to formally ensure $\dot{d}>0$, it is of minor importance in practice. Fatigue cracks at sub-critical loads are driven by a fatigue variable which is ever-increasing anyway.

\subsection{Acceleration methods for saving computational time} 
\label{sec:cycles}

Reducing computational time is crucial in cyclic phase-field fatigue simulations, especially, if elastic-plastic material models are involved. Not only for HCF simulations, cycle-by-cycle simulations are not feasible for components of practical relevance. The models mentioned here adress this problem mainly in two ways: Through representative loads (\textsc{Schreiber} et al. \cite{schreiber_phase_2020b}, \textsc{Seiler} et al. \cite{seiler_efficient_2020}) and through the cycle jump method (\textsc{Sele\u{s}} et al. \cite{seles_numerical_2020}, \textsc{Loew} et al. \cite{loew_fatigue_2020}, \textsc{Haveroth} et al. \cite{haveroth_nonisothermal_2020}).

\begin{figure} [h] 
	\def\svgwidth{\linewidth}\small{
		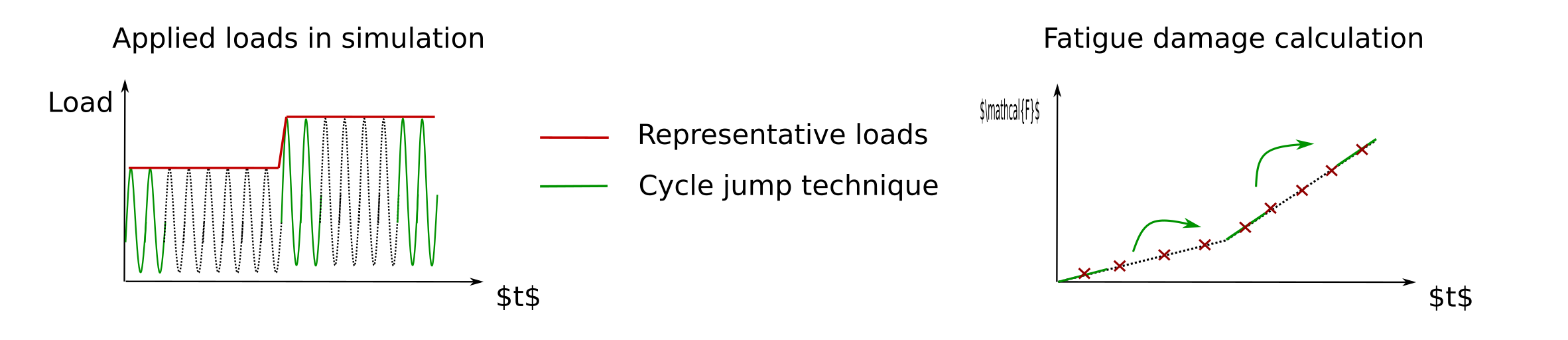}
	\caption{ Comparison of acceleration techniques: Representative loads and cycle jump method. Depiction of applied loads is based on \cite{schreiber_phase_2020b}. }
	\label{fig:acc}
\end{figure}

The latter is a general acceleration concept described by \textsc{Cojacaru} and \textsc{Karlsson} \cite{cojocaru_simple_2006}. As shown in Figure \ref{fig:acc} in green, a few cycles are simulated explicitly before the variables of interest -- the fatigue variable, plastic hardening variables etc. -- are extrapolated over a certain number of cycles. Then again follow properly simulated cycles. One difficulty is the choice of an appropriate jump size as a compromise between simulation time and accuracy, especially considering the often sudden nature of crack evolution.

Simulations with representative loads, on the other hand, are controlled by continuous fatigue instead of continuous time. As shown in Figure \ref{fig:acc} in red, not a single cycle is simulated explicitly. Instead, the load applied is a representative load, usually some sort of envelope curve of the real load function. The lack of information due to this simplification is compensated by assumptions which are mostly based on empirical fatigue concepts (see also Section \ref{sec:fatvar}). This can be an assumption of the stress-strain behaviour and the amount of damage depending on the area inside the stress-strain hysteresis (\textsc{Seiler} et al. \cite{seiler_efficient_2020}) or the damaging effect of load cycles according to their stress amplitude (\textsc{Schreiber} et al. \cite{schreiber_phase_2020b}), completed by cyclic material data such as \textsc{Wöhler} curves. In this way, the damage contribution of each cycle can be calculated from the stress-strain state at the representative load, possibly complemented with information regarding the load such as the ratio $R$ between maximum and minimum load. The choice of an appropriate representative load is always based on assumptions, such as that the most intensive crack driving state at the critical crack front happens during maximum load etc. Especially in case of variable amplitudes, several load levels might be necessary (e.\,g. maximum and minimum load) in order to quantify the damage contribution of that load cycle. This method has the greatest accelerating effect in the case of at least sectionwise constant load amplitudes, since, in that case, several load cycles can be combined in one increment if crack growth rates are small. A new increment is only necessary when significant crack growth has happened and changed the strange state in the specimen.

Lastly, \textsc{Lo} et al. \cite{lo_phasefield_2019} use an entirely different strategy, where they do not extrapolate fatigue damage $\calF$, but directly work with crack propagation rates fitted to \textsc{paris} curves. { For a review on acceleration techniques in a more general sense, including discretisation techniques, see \cite{cui_applications_2023}.}

\subsection{Additional stress terms} 
\label{sec:stress}

While in most models the stress is defined as $ \sig(\eps) = \diffp{W_\mathrm{el}}{\eps} $, \textsc{Schreiber} et al. \cite{schreiber_phase_2020b} and
\textsc{Haveroth} et al. \cite{haveroth_nonisothermal_2020} introduce additional stress terms. The reason for this is lies in their definition of the free energy density $\psi$, which includes -- in contrast to the definition used in Section \ref{sec:framework} -- fracture and fatigue terms, e.\,g. in \textsc{Schreiber} et al.
\begin{equation}
	\psi = g(d)\, \psie_+ + \psie_- + \Gc\gamma +  \gfat(d)H(\calF).
\end{equation}
Evaluating the \textsc{Clausius-Duhem} inequality $\sigma : \dot{\eps} - \dot{\psi}  \geq 0$ yields
\begin{equation}
	\underbrace{\sigma : \dot{\eps} - \diffp{\psi}{\eps}\dot{\eps} - \diffp{\psi}{\calF}\dot{\calF} }_{\text{(a)}} \underbrace{- \diffp{\psi}{d}\dot{d} - \diffp{\psi}{\nabla d}\left(\nabla d\right)^{\cdot} }_{\text{(b)}} \geq 0. 
\end{equation}
Supposing the common dependency $\calF(\eps)$, the stress is defined from $\text{(a)} \stackrel{!}{=} 0$ as
\begin{equation}
	\sigma = \diffp{\psi}{\eps} + \diffp{\psi}{\calF}\diffp{\calF}{\eps},
\end{equation}
which is in this case
\begin{equation}
	\sig = g(d) \mathbb{C} \eps + g(d)qb\langle \calF-\calF_{\min}\rangle^{b-1} \diffp{\calF}{\eps}.
\end{equation}
Term (b) yields 
\begin{equation}
	- \diffv{\psi}{d}\dot{d}\geq0,
\end{equation}
leading to the phase-field evolution equation.

Additional stress terms entail the issue of physical interpretation of those terms. \textsc{Schreiber} et al. interpret them as microscopic stresses.
However, even with this widespread extended definition of the free energy density $\psi$, additional stress terms are usually avoided by assuming $\calF$ to be constant in time for the considered time step, i.\,e. independent of $\eps$. This assumption is valid considering that $\calF$ changes on a large time scale (over the course of several load cycles) compared to e.\,g. the strain oscillating in each load cycle. 

\subsection{Range of application}

Finally, we want to give a list of models addressing certain types of scenarios and problems in simulation.

\subsubsection*{Material class}

While most models published are designed for metals, a few other material classes are adressed as well:
\begin{itemize}
	\item Elastomers: \textsc{Loew} et al. \cite{loew_fatigue_2020} with rate-dependent behaviour and a large strain setting, {\textsc{Yin} and \textsc{Kaliske} \cite{yin_fatigue_2023} with \textsc{Neo-Hooke} elastic material model}
	\item Concrete and rock: \textsc{Schröder} et al. {\color{blue}\cite{schroder_phasefield_2022,pise_phenomenological_2023}} with a \textsc{Drucker-Prager} yield criterion and unsymmetric tension-compression behaviour.
	\item Piezoelectric solids: \textsc{Tan} et al. \cite{tan_phase_2022} with coupling to electric field.
\end{itemize}

\subsubsection*{Loading}

For ductile materials like metals, high loading amplitudes (LCF) cause plasticity around the crack tip, therefore calling for an elastic-plastic material model. For low loading amplitudes (HCF) and brittle materials, an elastic model is sufficient.

\begin{itemize}
	\item Elastic: \textsc{Carrara} et al. \cite{carrara_framework_2019}, \textsc{Grossmann-Ponemon} et al. \cite{grossman-ponemon_phasefield_2022}, \textsc{Hasan} and \textsc{Baxevanis} \cite{hasan_phasefield_2021}, \textsc{Amendola} et al. \cite{amendola_thermomechanics_2016}, \textsc{Schreiber} et al. \cite{schreiber_phase_2020b}, \textsc{Lo} et al. \cite{lo_phasefield_2019} 
	\item Elastic-plastic: \textsc{Aygün} et al. \cite{aygun_coupling_2021a}, \textsc{Sele\u{s}} et al. \cite{seles_numerical_2020}, \textsc{Ulloa} et al. \cite{ulloa_phasefield_2021}, \textsc{Khalil} et al. \cite{khalil_phasefield_2021}, \textsc{Haveroth} et al. \cite{haveroth_nonisothermal_2020}
\end{itemize}

\subsubsection*{Observed phenomena and challenges}

\begin{itemize}
	\item Material behaviour dependent on deformation rate: \textsc{Loew} et al. \cite{loew_fatigue_2020}, \textsc{Haveroth} et al. \cite{haveroth_nonisothermal_2020}
	\item \textsc{Bauschinger} effect (kinematic hardening): \textsc{Aygün} et al. \cite{aygun_coupling_2021a}, \textsc{Sele\u{s}} et al. \cite{seles_numerical_2020}, \textsc{Ulloa} et al. \cite{ulloa_phasefield_2021}, \textsc{Khalil} et al. \cite{khalil_phasefield_2021}
	\item Ratchetting: \textsc{Ulloa} et al. \cite{ulloa_phasefield_2021}
	\item Temperature-dependent fatigue behaviour: \textsc{Amendola} et al. \cite{amendola_thermomechanics_2016}, \textsc{Haveroth} et al. \cite{haveroth_nonisothermal_2020}, \textsc{Yan} et al. \cite{yan_simulating_2022}
	\item Acceleration techniques for computational time: \textsc{Seiler} et al. \cite{seiler_efficient_2020}, \textsc{Sele\u{s}} et al. \cite{seles_numerical_2020}, \textsc{Schreiber} et al. \cite{schreiber_phase_2020b}, \textsc{Loew} et al. \cite{loew_accelerating_2020}, \textsc{Haveroth} et al. \cite{haveroth_nonisothermal_2020}, \textsc{Lo} et al. \cite{lo_phasefield_2019} 
	\item Concentration-dependent material behaviour: \textsc{Ai} et al. \cite{ai_coupled_2022} implemented a coupled chemo-mechanical fatigue fracture model to simulate cracking in lithium-ion batteries. The phase-field fatigue part is equivalent to \textsc{Carrara} et al. \cite{carrara_framework_2019}
	\item { Crystal microstructure to be simulated with RVE: \textsc{Lucarini} et al. \cite{lucarini_fftbased_2023}}
\end{itemize}

\section{Conclusion} 

In recent years, many groups have adressed the issue of fatigue fracture with a large variety of phase-field models.
This paper puts the models published to date into a common variational framework. Based on that, the model structures and characteristics are compared. This paper is meant to provide a basis for both choosing a model type for a specific simulation task and for developing phase-field models further.

Similarities and differences between the models are discussed. Thereby, two main model classes based on the model structure are identified: Firstly, $A$-type models that degrade the fracture toughness gradually in order to describe the continuous weakening of the material due to cyclic loading. And secondly, the $B$-type models characterised by an additional crack driving force compared to the static models, which allows the fatigue crack to propagate at the low fatigue loads. A numerical study shows that both model types actually suffer from fundamental problems regarding the regularised crack profile: While the $A$-type models degradation of the regularisation term in the phase-field evolution equation leads to narrower crack profiles compared to static cracks, $B$-type models show an unintended broadening of the crack profile due to the direct link between the distribution of the fatigue variable and the final crack profile.
Eventually, the choice between both model types should follow the preferred physical explanation of the incorporation of fatigue into the phase-field structure: While some might find a weakening of the material, associated with a decrease of total energy of the system, more plausible, others might prefer an additional fatigue energy contribution.

The second-most important modelling choice is the fatigue variable itself. Most groups choose the accumulated strain energy density as a fatigue measure. Not only is this quantity easily accessible, but also its significance as a measure of stored energy available for the forming of new crack surface straightforward. However, some models use empirical fatigue concepts instead. These incorporate additional cyclic material data in the calculation. The empirical assumptions inherent to the concepts actually allow for an acceleration scheme of the fatigue simulation. Alternatively, cycle jump concepts are widely used.

Essential for the choice of model are the material and the loading conditions. While most models are meant for metals, some also exist for other material classes. In case of low cycle fatigue with high loading amplitudes, elastic-plastic material models are to be favored due to their ability to model the significant plasticity at the crack tip. Elastic-plastic phase-field models differ in their way of incorporating plasticity in the fatigue variable.
By now, most models are able to reproduce typical phenomena observed in cyclic fracture experiments: \textsc{Wöhler} curves describing the lifetime of components as well as \textsc{Paris} curves for the crack propagation rates can be reproduced. Mean load effects are captured by some models.

Still, the simulation of fatigue cracks remains a challenging task, not only with the phase-field method. All models studied here are phenomenological and {(but one)} macroscopic. It is up to future works to develop models approaching the fatigue phenomenon from a more physical point of view, which always has to be -- at least in part -- microscopic. Multiscale models have not been used yet due to the immense computational power required for phase-field fatigue simulations. This is due to required fineness of meshes for phase-field simulations in general and, on the other hand, the high number of load cycles to be simulated inherent to cyclic loads. Especially for 3D simulations and elastic-plastic material behaviour, this problem sets the limits for simulations today.

\section*{Acknowledgements}

This work was supported by the Deutsche Forschungsgemeinschaft (DFG) via the project \textit{Experimental analysis and phase-field modelling of the interaction between plastic zone and fatigue crack growth in ductile materials under complex loading} (grant number KA 3309/12-1). The authors are grateful to the Centre for Information Services and High Performance Computing (ZIH) of TU Dresden for providing its facilities for high throughput calculations. The authors thank Franz Dammaß for comprehensive discussions and comments on the topic.

\section*{Highlights} 

\begin{enumerate}
	\item {Presentation of most existing phase-field fatigue models in common framework for the first time}
	\item Categorisation of the models in mainly two classes according to their structure
	\item Numerical comparison of the two model classes
	\item Discussion of incorporation of plasticity and acceleration techniques
\end{enumerate}

\appendix

\section{Derivation of plastic equations}
\label{ap:convexanalysis}

For the exemplary plastic dissipation potential 
\begin{equation}
	\phi^\mathrm{p}(\dot{\epsp},\dot{\te{\alpha}}) = \sigma^\mathrm{y} ||\dot{\epsp}|| + \frac{b}{2} \left( \dot{\epsp} + \dot{\te{\alpha}} \right)^2, \quad \sigy > 0
\end{equation}
the plastic equations are to be derived. From \textsc{Biot}'s equation (\ref{eq:biot}) follows for the plastic conjugate variables 
\begin{equation} \label{eq:chiA}
	\sig = \diffp{\phi^\mathrm{p}}{\dot{\epsp}} \quad \text{and} \quad \te{\chi} = \diffp{\phi^\mathrm{p}}{\dot{\alpha}} = b\,(\dot{\epsp} + \dot{\alpha}).
\end{equation}
For $\underline{||\dot{\epsp}|| \neq 0}$ the stress is
\begin{equation} \label{eq:sigA}
	\sig = \sigy \frac{\dot{\epsp}}{||\dot{\epsp}||} + b\,(\dot{\epsp}+\dot{\te{\alpha}}).
\end{equation}
From the difference (\ref{eq:sigA})-(\ref{eq:chiA}) we get 
\begin{equation}
	\sig-\te{\chi} = \sigy \frac{\dot{\epsp}}{||\dot{\epsp}||} \quad \text{and} \quad ||\sig-\te{\chi}||= \sigy.
\end{equation}
Defining $\lambda = ||\dot{\epsp}||$ and $f^\mathrm{p} = ||\sig-\te{\chi}||-\sigy$ we obtain
\begin{align}
	& \dot{\epsp} =\lambda \frac{\sig-\te{\chi}}{\sigy} = \lambda \diffp{f^\mathrm{p}}{\sig} \label{eq:epsneq0_1}\\
	& \lambda = ||\dot{\epsp}|| \geq0 \\
	& f^\mathrm{p} = ||\sig-\te{\chi}|| - \sigy = 0 \\
	& \lambda f^\mathrm{p} = 0. \label{eq:epsneq0_4}
\end{align}
For the case $\underline{||\dot{\epsp}|| = 0}$ the derivative $\sig = \diffp{\phi^\mathrm{p}}{\dot{\epsp}}$, especially the problematic term $\diffp{}{\dot{\epsp}} ||\dot{\epsp}||$ has to yet to be defined. Here, the fact that the absolute value $||\dot{\epsp}||$ is a convex function can be exploited. Using convex analysis, its derivative
is defined as
\begin{equation}
	\diffp{}{\dot{\epsp}} ||\dot{\epsp}|| \bigg\vert_{\dot{\epsp}=0} = \mu \te{n} \quad \text{with } 0\leq\mu\leq1, ||\te{n}||=1.
\end{equation}
Vividly speaking, the derivative of the absolute value function at its kink is defined with arbitrary direction. With
\begin{equation}
	\sig = \sigy \mu \te{n} + b\,(\dot{\epsp} + \dot{\te{\alpha}})
\end{equation}
and (\ref{eq:chiA}) follows
\begin{equation}
	\sig-\te{\chi} = \sigy\mu\te{n} \quad \text{and} \quad \frac{||\sig-\te{\chi}||}{\sigy} = \mu.
\end{equation}
From $0\leq\mu\leq1$ follows 
\begin{align}
	&	||\dot{\epsp}|| = 0 \label{eq:epsis0_1}\\
	& 	\lambda = ||\dot{\epsp}|| = 0 \\
	&	||\sig-\te{\chi}|| - \sigy \leq 0 \rightarrow f^\mathrm{p} \leq 0 \\
	& \lambda f^\mathrm{p} =0. \label{eq:epsis0_4}
\end{align}
From the sets of equations for the two cases (\ref{eq:epsneq0_1})..(\ref{eq:epsneq0_4}) and (\ref{eq:epsis0_1})..(\ref{eq:epsis0_4}) follow for \underline{all $||\dot{\epsp}||$} 
the evolution equation for the plastic strain and the KKT conditions 
\begin{equation}
	\dot{\epsp} = \lambda \diffp{f^\mathrm{p}}{\sig} \quad \text{and} \quad \lambda\geq0, f^\mathrm{p}\leq0, \lambda f^\mathrm{p}=0. 
\end{equation} 
The time derivative of (\ref{eq:epsis0_4}) for the case $f^\mathrm{p}=0$ yields the consistency condition $\lambda\dot{f}^\mathrm{p}=0$.

\section{Derivation of plastic model equations via dissipation potential }
\label{ap:maxdiss}

From the \textsc{Clausius-Duhem} inequality 
\begin{equation}
	\sig:\dot{\eps} - \diffp{\psi}{\eps}:\dot{\eps} - \diffp{\psi}{\epsp}:\dot{\epsp} - \diffp{\psi}{\te{\alpha}}:\dot{\te{\alpha}} - \diffp{\psi}{\alpha}\dot{\alpha} \geq 0
\end{equation}
we can identify the plastic conjugate variables
\begin{equation}
	-\diffp{\psi}{\epsp} =:  \sig, \quad -\diffp{\psi}{\te{\alpha}}  =: \te{\chi}, \quad -\diffp{\psi}{\alpha} =: p .
\end{equation}
A yield function
\begin{equation}
	f^\mathrm{p}(\sig,\chi,p;d) \Def \sqrt{\frac{3}{2}||\dev(\sig)-\dev({\te{\chi}})||^2} - \sigma^\mathrm{y} + p   
\end{equation}
is defined. The evolution equations for the internal variables can now be derived e.g. from the principle of maximum plastic dissipation
\begin{equation}
	\phi^\mathrm{p} = \sup_{\sig,\boldsymbol{\chi},p,\lambda,z} \hat{\phi}^\mathrm{p} = \sup_{\sig,\boldsymbol{\chi},p,\lambda,z} \left\{ \sig:\epsp + \te{\chi}:\dot{\te{\alpha}} + p\, \dot{\alpha}  - \lambda(f^\mathrm{p}(\sig,\chi,p;d) + z^2) \right\},
\end{equation}
constrained by the yield function $f^\mathrm{p}$ using a \textsc{Lagrange} multiplier $\lambda$ and a slack variable $z$. The supremum requires the partial derivatives of $\hat{\phi}^\mathrm{p}$ with respect to $\sig,\te{\chi},p,\lambda,z$ to be 0 as well as $\diffp{^2\hat{\phi}^\mathrm{p}}{z^2}\leq 0$. This yields the flow rule and the hardening laws 
\begin{equation}
	\dot{\epsp} = \lambda\diffp{f^\mathrm{p}}{\sig} = \lambda \ve{n}^\mathrm{p}
	, \quad   \dot{\te{\alpha}} = \lambda\diffp{f^\mathrm{p}}{\te{\chi}}= \lambda \ve{n}^\mathrm{p} \quad \text{and} \quad \dot{\alpha} = \lambda\diffp{f^\mathrm{p}}{p}
\end{equation}
with direction tensor $\ve{n}^\mathrm{p}$, as well as the KKT conditions
\begin{equation}
	f^\mathrm{p}\leq0, \quad \lambda\geq0 \quad \text{and} \quad f^\mathrm{p}\lambda =0.
\end{equation}
and, following for $f^\mathrm{p}=0$, the consistency condition $\lambda\dot{f}^\mathrm{p}=0$.

\section{Phase-field equation with penalty approach }
\label{ap:penalty}

An alternative penalisation approach to ensure $\dot{d}\geq0$ is proposed by \cite{gerasimov_penalization_2019}. A modified energy functional 
\begin{equation}
	\tilde{\calE}(\eps,d,\nabla d,\dot{d},\qal,\dot{\te{q}}_{\alpha};\calF) = \calE + \frac{\lambda^\infty}{2} \inte{\calB}{\langle \dot{d} \rangle^2_-}{v}
\end{equation}
is introduced. The penalty term with penalty parameter $\lambda^\infty$ and $\langle x\rangle_- \Def \min(0,x)$ yields the evolution equation
\begin{equation} 
	\eta \dot{d} = \Gc{\color{blue}h(\calF)}\left(\ell\Delta d - \frac{d}{\ell}\right) {\color{blue} + \Gc\ell\nabla d\nabla h(\calF)} 
	- \lambda^\infty\langle  \dot{d}\rangle_-  -g'(d)\underbrace{\left(\psie_+(\epse) + \psip(\alpha,\te{\alpha})   {\color{red} + H(\calF)} + \Delta^\mathrm{p}\right)}_{\mathcal{H}} .
\end{equation}

\bibliography{mybibfile}[url=false]	

\end{document}